\documentclass[apsprl,twocolumn,superscriptaddress]{revtex4-2}

\usepackage{graphicx,subfigure}
\usepackage{changes}
\usepackage{amsmath,amssymb}
\usepackage{mathtools}
\usepackage{multirow}
\usepackage{textcomp}
\usepackage{float}
\usepackage{xcolor}
\usepackage{titletoc}
\usepackage{ulem}
\usepackage{dsfont}
\usepackage{siunitx}
\DeclareSIUnit\gauss{G}
\usepackage{bibunits}
\usepackage{import}
\usepackage{braket}
\usepackage{nicefrac}
\usepackage{comment}
\usepackage{bm}
\usepackage{bbm}
\usepackage[english]{babel}
\usepackage{hyperref}
\hypersetup{colorlinks=true,linkcolor=blue,citecolor=blue,breaklinks}

\newcommand{\expect}[1]{\ensuremath{\left\langle{#1}\right\rangle}}

\newcommand{\micro}[1]{\ensuremath{\mu\mathrm{#1}}}

\renewcommand{\micro}[1]{\ensuremath \mu\mathrm{#1}}

\renewcommand{\vec}[1]{\ensuremath{\mathbf{#1}}}

\let\oldsfdefault\sfdefault
\usepackage[scaled=0.92]{helvet}
\renewcommand{\sfdefault}{\oldsfdefault}

\definechangesauthor[name=Immanuel Bloch, color=red]{ib}

\usepackage{layouts}

\defaultbibliographystyle{apsrev4-2}

\begin{document}

\begin{bibunit}

\title{Local readout and control of current and kinetic energy operators in optical lattices}

\author{Alexander Impertro}
\author{Simon Karch}
\author{Julian F. Wienand}
\author{SeungJung~Huh}
\author{Christian Schweizer}
\author{Immanuel Bloch}
\author{Monika Aidelsburger}
    \affiliation{Fakult\"{a}t f\"{u}r Physik, Ludwig-Maximilians-Universit\"{a}t, 80799 Munich, Germany}
    \affiliation{Max-Planck-Institut f\"{u}r Quantenoptik, 85748 Garching, Germany}
    \affiliation{Munich Center for Quantum Science and Technology (MCQST), 80799 Munich, Germany}

\date{\today}


\begin{abstract}

Quantum gas microscopes have revolutionized quantum simulations with ultracold atoms, allowing to measure local observables and snapshots of quantum states. However, measurements so far were mostly carried out in the occupation basis. Here, we demonstrate how all kinetic operators, such as kinetic energy or current operators, can be measured and manipulated with single bond resolution. Beyond simple expectation values of these observables, the single-shot measurements allow to access full counting statistics and complex correlation functions. Our work paves the way for the implementation of efficient quantum state tomography and hybrid quantum computing protocols for itinerant particles on a lattice. In addition, we demonstrate how site-resolved programmable potentials enable a spatially-selective, parallel readout in different bases as well as the engineering of arbitrary initial states.

\end{abstract}
\maketitle


Analog quantum simulators offer a promising route towards practical quantum advantage, being robust in the measured observables and against small imperfections~\cite{cirac_goals_2012, preskill_quantum_2018, trivedi_quantum_2022, daley_practical_2022}. Among those, neutral atoms in optical lattices are ideal candidates for simulating a large variety of condensed matter models~\cite{gross_quantum_2017,halimeh_cold-atom_2023}, providing access to single-atom and single-site resolved detection of (non-local) correlation functions and counting statistics through quantum gas microscopy (QGM)~\cite{bakr_quantum_2009,sherson_single-atom-resolved_2010,gross_quantum_2021}. However, most measurements in these platforms so far were carried out in the occupation basis, limiting the range of state-preparation and read-out protocols that can be implemented. A measurement of the current operator would for example aid in the study of non-equilibrium dynamics by allowing to probe information scrambling through bond-resolved local currents and off-diagonal correlations~\cite{swingle_measuring_2016, bohrdt_scrambling_2017}, as well as in the simulation of interacting topological phases that host equilibrium currents and current vortices~\cite{piraud_vortex_2015, wang_measurable_2022}. Furthermore, a measurement in a complete basis would enable the implementation of Hamiltonian learning, which is a promising approach for benchmarking analog quantum simulators~\cite{wiebe_hamiltonian_2014,wang_experimental_2017,carrasco_theoretical_2021,yu_robust_2023}.

\begin{figure}[!htb]
    \centering
    \includegraphics[width=\columnwidth]{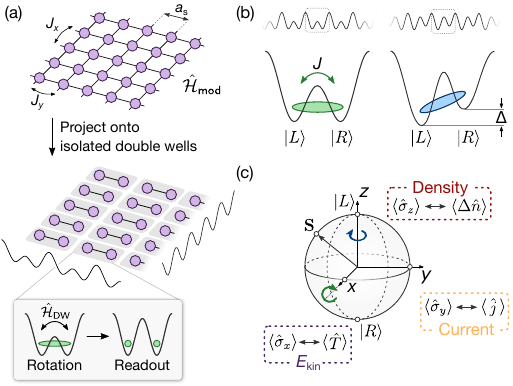}
    \caption{\textbf{Programmable rotations using double-well potentials.} (a) A system evolving under a lattice Hamiltonian $\hat{\mathcal{H}}_\mathrm{mod}$ with coherent tunneling is projected onto isolated double wells (DWs), which are created using a bichromatic optical superlattice. Simultaneously in all DWs, a programmable operation is applied according to $\hat{\mathcal{H}}_\mathrm{DW}$ [defined in the main text and panel (b)] to locally rotate the measurement basis. The occupation in the DWs is then frozen and read out with local resolution. (b) Atomic operations in a DW potential, forming a two-level system out of the states $\ket{L}$ and $\ket{R}$. A symmetric, coupled DW with coupling strength $J$ realizes an $X$ rotation (left). A strongly tilted DW with tilt $\Delta$ implements a $Z$ rotation (right). (c) The rotations are used to map the current $\hat{j}$ and the kinetic energy operator $\hat{T}$ onto density imbalance $\Delta\hat{n}$ for local read-out. With this, all components of the Bloch vector $\vec{S}=(\hat{\sigma}_x,\hat{\sigma}_y,\hat{\sigma}_z)$ are experimentally accessible.}
    \label{fig:experiment_scheme}
\end{figure}

In this Letter, we demonstrate how -- in addition to the density -- the kinetic energy and the current operators, or any linear combination of the two (\textit{kinetic operators}), can be measured and controlled with local resolution using optical superlattices. Optical superlattices enable parallel high-fidelity nearest-neighbor manipulations, which have been used to generate a large number of entangled atom pairs based on superexchange interactions as well as scalable entanglement~\cite{trotzky_time-resolved_2008,trotzky_controlling_2010,yang_cooling_2020,zhang_scalable_2023}. Here, we use superlattices to project a many-body system on a two-dimensional (2D) lattice onto isolated double wells (DWs), as depicted in Fig.~\ref{fig:experiment_scheme}a. The sites of the DW form a two-level system, where tunnel coupling $J$ and a potential energy difference $\Delta$ can be interpreted as Pauli-$X$ and $Z$ operations (Fig.~\ref{fig:experiment_scheme}b). This has been used in an earlier experiment to measure a spatially-averaged current operator~\cite{atala_observation_2014}. In this work we extend these ideas and combine them with local resolution and manipulation techniques. We further demonstrate arbitrary rotations in the DWs by combining $X$ and $Z$ operations (Fig.~\ref{fig:experiment_scheme}c). This is used to measure the current and kinetic-energy operator with local resolution and in a single experimental realization, providing access to correlations and counting statistics~\cite{kesler_single-site-resolved_2014}. We further apply site-resolved programmable potentials to perform spatially-selective basis rotations as well as coherent manipulations to engineer states with flexible density and phase patterns, including states with local coherent superpositions. Our technique is directly applicable to interacting quantum systems given that interactions can be switched off during the DW manipulations~\cite{kaufman_quantum_2016,su_observation_2023}. 

\begin{figure*}[t]
    \centering
    \includegraphics[width=\textwidth]{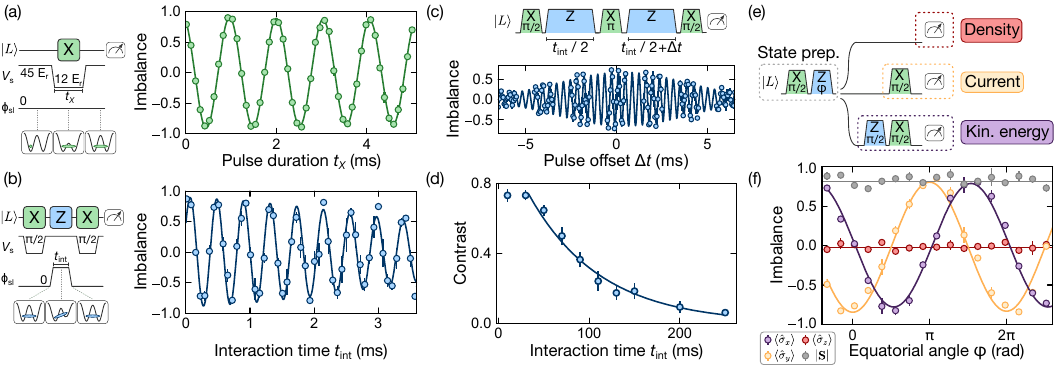}
    \caption{\textbf{Global double-well operations and rotations of the measurement basis.} (a) $X$ rotation starting with the DWs initialized in $\ket{L}$. $V_\mathrm{s}$ denotes the depth of the short lattice, which sets the tunnel coupling $J$ inside the well, and $\phi_\mathrm{sl}$ is the superlattice phase. The solid line is a fit to an exponentially damped sine, yielding an imbalance oscillation corresponding to a tunnel coupling of $J = h \times \SI{484.3(5)}{Hz}$. The data has been evaluated in a region of interest (ROI) spanning $28\times32$ sites, and the error bars are the standard error of the mean (s.e.m.) over 6 repetitions for each data point. (b) $Z$ rotations and Ramsey sequence. A $Z$ rotation is achieved by jumping the superlattice phase to around $\SI{0.1}{\radian}$ away from the symmetric point. The solid line is a fit to a numerical model accounting for on-site potential disorder (see SM for details). The evaluation ROI was $16\times 30$, and the error bars are the s.e.m. over 4 repetitions. (c) Spin-echo sequence. Example trace for $t_\mathrm{int}=\SI{30}{ms}$, recorded by varying the time offset $\Delta t$ in the second $Z$ rotation. The solid line is a fit of a sine with a Gaussian envelope function. The evaluation ROI was $40\times40$ sites, and each data point is averaged over 5 repetitions. (d) Measurement of the $T_2$ time using the spin-echo sequence. For each data point, the imbalance contrast was evaluated by varying the pulse offset and fitting the resulting imbalance oscillation. The solid line is an exponential fit excluding the first point, from which we determine a $T_2$ time of $\SI{113(10)}{ms}$ (defined as the time where the contrast has decreased to a fraction of $1/e$ of the first data point). The error bars are the standard errors of the fit, other evaluation details as in panel (c). (e) Scheme to determine density, current and kinetic energy for the states lying on the equator of the Bloch sphere. (f) Measurement result as a function of the equatorial angle $\varphi$. The solid lines for the current (yellow) and the kinetic energy (purple) are fits to a sine. The solid line for the density is a fit to a constant function. The gray data points show the length of the Bloch vector $|\Vec{S}| = \sqrt{\expect{\hat{\sigma}_x}^2+\expect{\hat{\sigma}_y}^2+\expect{\hat{\sigma}_z}^2}$, which is $0.81(6)$ on average (solid gray line). The zero of the horizontal axis has been calibrated on the first minimum of the $\hat{\sigma}_y$ trace. The evaluation ROI was $18\times36$ sites, and the error bars are the s.e.m. over 3 repetitions.}
    \label{fig:global_ops}
\end{figure*}

\textit{Experimental scheme.} We create an optical superlattice potential by superimposing two standing waves differing by a factor of two in wavelength, which results in a potential of the form $V_\mathrm{sl}(x) = V_\mathrm{s} \cos^2\left(k_\mathrm{s}x\right) + V_\mathrm{l} \cos^2\left(k_\mathrm{l}x + \phi_\mathrm{sl}/2\right)$~\cite{folling_direct_2007}.
Here $V_\mathrm{s(l)}$ is the lattice depth, $k_\mathrm{s(l)}=\pi/a_\mathrm{s(l)}$ the wave vector of the short(long)-period lattice [with $a_\mathrm{s(l)}$ being the lattice constant of the short(long)-period lattice, where $a_\mathrm{s}=\SI{383.5}{nm}=a_\mathrm{l}/2$] and $\phi_\mathrm{sl}$ the superlattice phase. This realizes a periodic array of tunable DWs~(see~Fig.~\ref{fig:experiment_scheme}b). For a deep long-period lattice, the inter-DW coupling is negligible and the states $\ket{L}$ and $\ket{R}$ in each DW form a two-level system with the Hamiltonian $\hat{\mathcal{H}}_\mathrm{DW} = -J\hat{\sigma}_x-\frac{\Delta}{2}\hat{\sigma}_z$, where $J$ is the coupling matrix element between the two wells, $\Delta$ is the energy difference between the states and $\hat{\sigma}_i$ ($i=\{x,y,z\}$) are the Pauli operators. For symmetric DWs ($\Delta=0$), an $X$ rotation is realized (Fig.~\ref{fig:experiment_scheme}b, left), while a strongly tilted, decoupled DW ($\Delta \gg J$) implements a $Z$ rotation (Fig.~\ref{fig:experiment_scheme}b, right).

\begin{figure*}[t]
    \centering
    \includegraphics[width=\textwidth]{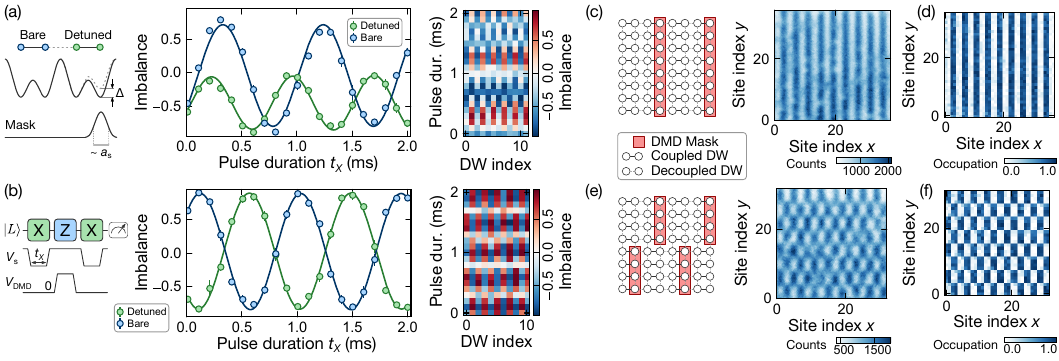}
    \caption{\textbf{Local programmable double-well operations.} (a) Locally detuned $X$ rotations. Using a digital micromirror device (DMD), we project a repulsive potential that locally tilts every other DW in a one-dimensional (1D) superlattice potential. The DMD mask consists of bright stripes with a width of $0.6\,a_\mathrm{s}$, which is broadened by the point-spread function to $\sim1-2\,a_\mathrm{s}$. The solid lines are fits to a sine, yielding a differential tilt of $\Delta=h\times\SI{797(13)}{Hz}$. Evaluation ROI: $24\times24$ sites, and the spatially-resolved plot is averaged over the perpendicular direction. Error bars show the s.e.m. over 3 repetitions, and are sometimes smaller than the marker size. (b) Local $Z$ rotations, implemented by tilting every other DW along a 1D chain. The $Z$ pulse area was chosen to be around $\pi$, which is visualized by scanning the duration of the first $X$ pulse, resulting in out-of-phase imbalance oscillations. The solid lines are fits to a sine, yielding a relative phase shift of $1.04(2)\,\pi$. Evaluation details as in (a). (c) State engineering using locally-detuned DW oscillations with the programmed mask shown on the left. The panel on the right shows a single fluorescence image of the resulting $\ket{110011...}$ state. (d) Averaged occupation for the state in (c), computed from 30 repetitions. (e) State engineering in two spatial dimensions (mask on the left), resulting in a checkerboard-like state of $4\times2$ site blocks. The panel on the right shows a single fluorescence image. (f) Averaged occupation for the state in (e), computed from 30 repetitions.}
    \label{fig:local_ops}
\end{figure*}

Local site-resolved densities $\expect{\hat{n}_{L,R}}$ are directly accessible in experiments. Here, $\hat{n}_{i}=\hat{a}^\dagger_i\hat{a}^{\phantom\dagger}_i$ is the bosonic number operator and $\hat{a}^\dagger_i$ the bosonic creation operator for the state $i$ ($i=\{L,R\}$), respectively. A measurement in the $Z$-basis then corresponds to probing the density difference within one DW, i.e., $\Delta \hat{n} = \hat{n}_L-\hat{n}_R = \hat{\sigma}_z$. In contrast, the local current between the two wells can be defined for a Hubbard model with real-valued tunneling $J$ as ${\hat{j} = i J \left( \hat{a}^\dagger_R\hat{a}^{\phantom\dagger}_L - \hat{a}^\dagger_L\hat{a}^{\phantom\dagger}_R \right) = J\hat{\sigma}_y}$~\cite{atala_observation_2014,kesler_single-site-resolved_2014}, requiring a measurement in the $Y$ basis. Similarly, a measurement in the $X$ basis gives the expectation value of the kinetic energy operator $\hat{T} = -J \left( \hat{a}^\dagger_R\hat{a}^{\phantom\dagger}_L + \hat{a}^\dagger_L\hat{a}^{\phantom\dagger}_R \right) = -J \hat{\sigma}_x$. 

To measure the kinetic operators experimentally, we use the DW dynamics to map them onto density imbalance $\Delta \hat{n}$. As illustrated in Fig.~\ref{fig:experiment_scheme}c, the current can be mapped onto the density via an $X_{\pi / 2}$ rotation, and the kinetic energy by concatenating a $Z_{\pi / 2}$ rotation with an $X_{\pi / 2}$ rotation (cf. also Fig.~\ref{fig:global_ops}e). In particular, the density difference after an $X_{\pi / 2}$ rotation can be written as $\expect{\Delta \hat{n}} = - \langle\hat{j}\rangle / J$, and after a $(Z_{\pi / 2}, X_{\pi / 2})$ sequence as $\expect{\Delta \hat{n}} = - \langle\hat{T}\rangle / J$ (see SM for a derivation). This realizes a single-shot readout of the density, kinetic energy and current with local resolution.

\textit{Results.} The experimental sequence starts by loading ultracold cesium atoms into a 2D optical lattice potential that consists of a superlattice in the $x$ direction as well as a single-color lattice in the $y$ direction, leading to an array of DWs along $x$ (see SM and Refs.~\cite{impertro_unsupervised_2023,wienand_emergence_2023} for details). We prepare a product state where all DWs are initialized with one particle in $\ket{L}$, which has a maximum imbalance $\mathcal{I}=(n_L-n_R)/(n_L+n_R)$ of typically $0.93(4)$ (see SM for details on the initial-state quality). Both short lattices~($x,y$) are at a depth of $45\,E_\mathrm{r,s}$ and the long lattice in~$x$ is at a depth of $45\,E_\mathrm{r,l}$ [$E_\mathrm{r,s(l)}=h^2/(8 m a_\mathrm{s(l)}^2)$ is the recoil energy of the short(long) lattice, $h$ is Planck's constant and $m$ is the atomic mass of cesium]. 

We implement an $X$ rotation by setting the superlattice phase to zero (symmetric DWs), quenching the short lattice depth along $x$ to $12\,E_\mathrm{r}$ in $\SI{200}{\micro\second}$ and letting the state evolve for a controllable duration. This leads to high-contrast imbalance oscillations as shown in Fig.~\ref{fig:global_ops}a. From the fit we determine a tunnel coupling of ${J = h \times \SI{484.3(5)}{Hz}}$, corresponding to a $\pi$ time of ${t_\pi = \SI{449(3)}{\micro\second}}$, and a $1/e$ decay constant of ${\tau = \SI{57(13)}{\milli\second}}$. Using the decay envelope we estimate the fidelity of a single $X_{\pi}$ pulse as ${\mathcal{F} = \SI{99.2(2)}{\percent}}$ (see SM for details). This fidelity is mostly limited by spatially-inhomogeneous potential energy variations which detune the DWs locally and modify the oscillation frequency according to $f=\sqrt{4J^2 + \Delta^2} / h$.

Next, we demonstrate $Z$ rotations using a Ramsey sequence. After an $X_{\pi / 2}$ pulse, we jump the superlattice phase away from the symmetric configuration, causing the Bloch vector to rotate on the equator as the superposition of $\ket{L}$ and $\ket{R}$ time-evolves in the tilted DW potential (cf. Fig.~\ref{fig:experiment_scheme}b). This evolution is probed using a second $X_{\pi / 2}$ pulse, yielding oscillations that reveal the rotation of the state vector along the equator (Fig.~\ref{fig:global_ops}b). The oscillations correspond to a tilt of $\Delta=h\times\SI{2.406(5)}{kHz}$, and exhibit a damping that is consistent with an on-site white noise disorder of amplitude $W=h\times\SI{49(2)}{Hz}$ (see SM for details). The envelope can also be approximated by a single exponential, giving a $T_2^*$ time of $\SI{6(1)}{ms}$ ($1/e$ decay).

Here, dephasing occurs faster compared to $X$ rotations, as local potential variations modify the tilt linearly, in contrast to the quadratic correction for $X$ rotations. To cancel the dephasing due to static potential disorder, we employ a spin-echo sequence as shown in Fig.~\ref{fig:global_ops}c. We determine a $T_2$ time of $\SI{113(10)}{ms}$~(Fig.~\ref{fig:global_ops}d), corresponding to around 270 interaction times at the previously measured $\Delta$. The $T_2$ time is more than an order of magnitude larger than the $T_2^*$ decay of the Ramsey signal and confirms that the dephasing is dominated by static potential inhomogeneities (see SM for an estimation of the dynamic disorder).

We use the rotations introduced above to perform a global measurement in all three Pauli bases. As an example, we prepare equatorial states of the form ${(\ket{L}+e^{i\varphi}\ket{R})/\sqrt{2}}$ by concatenating a $X_{\pi / 2}$ pulse and a $Z$ pulse of variable duration to tune the equatorial angle $\varphi$. We then measure the density ($\hat{\sigma}_z$), the current operator ($\hat{\sigma}_y$) and the kinetic energy operator ($\hat{\sigma}_x$) (see Fig.~\ref{fig:global_ops}e). The result of this is shown in Fig.~\ref{fig:global_ops}f. As expected for equatorial states, $\langle\hat{\sigma}_z\rangle$ is zero and time-independent, while both the expectation values for the current and the kinetic energy show high-contrast oscillations with a relative phase shift of $\pi/2$. We also determine the length of the Bloch vector as ${|\Vec{S}| = \sqrt{\expect{\hat{\sigma}_x}^2+\expect{\hat{\sigma}_y}^2+\expect{\hat{\sigma}_z}^2}}$, yielding an average length of $0.81(6)$. This value is expected to be predominantly limited by the preparation of the equatorial states.

The operations above allow to measure the entire system in the same basis. In addition, using local control, we can simultaneously measure different parts of the system in different bases. This can for example be used to enhance the measurement sensitivity for metrology applications~\cite{shaw_multi-ensemble_2023}, as well as to access non-trivial correlators between current and kinetic energy. In particular, we employ a digital micromirror device (DMD) to project programmable repulsive potentials and locally tilt selected DWs. As an example, we perform global $X$ rotations and simultaneously tilt every other DW, as illustrated in Fig.~\ref{fig:local_ops}a. The tilted DWs can be seen oscillating at a higher frequency and at a smaller amplitude, as expected from detuned Rabi oscillations. The maximum possible tilt is limited by the available power of the DMD light as well as the resolution, where the latter causes light to spill over into adjacent sites and reduce the differential tilt. A spatially-resolved evaluation reveals that this manipulation is also possible in a parallel fashion across extended regions of the system (see panel on right-hand side of Fig.~\ref{fig:local_ops}a).

A second application lies in programming local $Z$ rotations using DMD-imprinted tilts. We leverage these for a Ramsey-like sequence as shown in Fig.~\ref{fig:local_ops}b. In particular, we locally apply $Z_{\pi}$ rotations on every other DW, imprinting a $\pi$ relative phase into the local initial state. Scanning the duration of the first $X$ pulse results in strong out-of-phase oscillations with a fitted phase shift of $\Phi = 1.04(2)\,\pi$. This can immediately be used to realize a simultaneous measurement of current and kinetic energy in different locations using a local $Z_{\pi/2}$ rotation, or more generally for constructing spatially more complex observables.

Besides changing the measurement basis, the local rotations can also be used for the precise coherent engineering of complex spatially-structured states. As an example, we use locally-detuned $X$ rotations, where the imprinted tilt is chosen such that a minimum in the detuned DW imbalance coincides with a maximum in the bare DWs (e.g. around $t_X=\SI{1.3}{ms}$ in Fig.~\ref{fig:local_ops}a, see SM for details on the sequence). Sitting at this point, we can coherently transfer the initial $\ket{101010...}$ state into $\ket{110011...}$. In Fig.~\ref{fig:local_ops}c, this is demonstrated using a DMD mask that is translationally invariant in the direction perpendicular to the DWs. The averaged occupation (Fig.~\ref{fig:local_ops}d) indicates a filling of $86(4)\,\%$ in the occupied, and $7(2)\,\%$ in the empty stripes. Within the error bar, the quality of the engineered state is equal to the initial state, suggesting a high preparation fidelity (see SM for the initial-state quality). Similarly, we can choose a DMD mask that has an alternating pattern in the direction perpendicular to the DWs. This realizes a checkerboard-like state made up of $4 \times 2$ site blocks~(Fig.~\ref{fig:local_ops}e,f) with similarly high average fillings of $84(5)\,\%$ in the occupied, and $8(3)\,\%$ in the empty blocks. The attainable preparation fidelities are mostly limited by the resolution of the DMD projection system as well as a correct alignment of the projected mask relative to the lattice. Note that we achieve considerable differential tilts and transfer fidelities despite our particularly small lattice spacing of $a_\mathrm{s}=\SI{383.5}{nm}$. At a larger lattice spacing or with a better resolution, this scheme can be expected to allow for even more robust high-fidelity operations. 

\begin{figure}[t]
    \centering
    \includegraphics[width=\columnwidth]{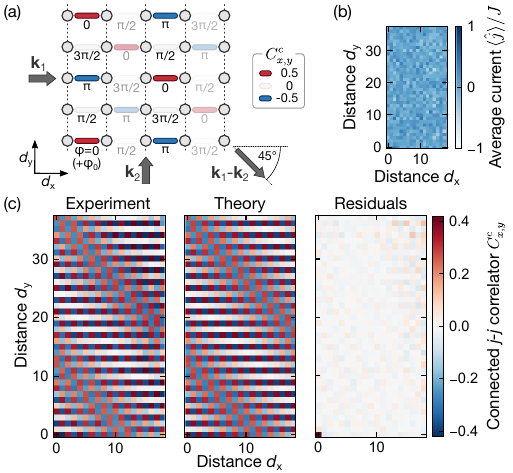}
    \caption{\textbf{Resolving spatially-varying complex tunnel couplings using current-current correlations.} (a) Scheme to generate spatially-varying tunnel coupling phases using a running-wave modulation created from the interference of two laser beams with wave vectors $\vec{k}_{1,2}$. The geometric arrangement results in a $\pi/2$ advancing phase pattern. The local phase $\varphi$ between sites can be revealed using a current measurement, as the local current is given by $\langle\hat{j}\rangle / J = \sin\varphi$. The color coding on the bonds illustrates the expected current-current-correlations. Due to the projection onto isolated DWs, only every second phase in the $x$ direction is accessible (light shading).  (b) Averaged current signal, computed from 50 realizations. Due to the random phase offset $\varphi_0$ of the running wave in every shot, the average current is featureless. (c) 2D connected current-current correlator revealing the underlying phase pattern. Experimental data from 50 averages (left), fit to a theoretical model (middle) and fit residuals (right). The fit yields an amplitude of $A=0.777(3)$, a common-mode angle between the laser-assisted tunneling beams and the lattice of $\theta_c=\SI{1.802(3)}{^\circ}$ as well as a relative angle deviation from $\SI{90}{^\circ}$ of $\theta_r=-\SI{0.558(8)}{^\circ}$, where the uncertainties are understood as the standard errors of the fit (see SM for details on the fit).}
    \label{fig:running_wave}
\end{figure}

Lastly, we would like to highlight that the single-shot and locally-resolved nature of the presented measurements furthermore enable the extraction of correlation functions. This is crucial, for example when spatial features are not stationary between individual shots. An illustrative example for this is found in laser-assisted tunneling schemes that are used to engineer artificial gauge fields with complex-valued tunnel couplings~\cite{aidelsburger_experimental_2011, miyake_realizing_2013, tai_microscopy_2017}. In our experiment, we employ a running-wave modulation that induces tunnel couplings with a complex phase factor increasing by about $\pi/2$ per bond in both spatial directions~(see Fig.~\ref{fig:running_wave}a). The ground states of isolated DWs subject to this modulation is $\ket{\psi} = (\ket{L} + e^{i\varphi}\ket{R})/\sqrt{2}$, where $\varphi$ is the local bond phase. In this state, the system has a homogeneous, constant density without any dependency on time and space, hindering a measurement of the local phase in the $Z$ basis. The expectation value of the current operator for this state is $\langle\hat{j}\rangle / J = \sin\varphi$, indicating that a current measurement can be performed to reveal this spatially-varying phase pattern. However, due to the running-wave, the global phase of this pattern varies randomly between experimental realizations, washing out any signal in the averaged current operator~(Fig.~\ref{fig:running_wave}b). Instead, we can detect the phase pattern by evaluating the connected 2D current-current correlation function ${C^c_{x,y} = \langle\hat{j}_{i,j} \hat{j}_{i+x,j+y}\rangle - \langle\hat{j}_{i,j}\rangle\langle\hat{j}_{i+x,j+y}\rangle}$, where $\hat{j}_{i,j}$ denotes the current on the DW at location $(i,j)$ in the 2D lattice. 

Fig.~\ref{fig:running_wave}c shows the experimentally measured 2D correlation function (left panel), which, focusing on small distances ($d_{x,y}\lesssim 4$), matches our expectation given the phase pattern in Fig.~\ref{fig:running_wave}a. For larger distances, we find Moiré fringes, which arise from a slight angular misalignment between the lattice base vectors and the running-wave modulation beams away from $\pi/2$. As can be seen in Fig.~\ref{fig:running_wave}c (middle and right panel), we find excellent agreement with a theoretical model that accounts for relative angles (see SM for further details on the fitting model). The measured correlator amplitude is around $78\%$ of the ideal value, which we believe to be mostly limited by a non-adiabatic ground state preparation.

\textit{Conclusion.} We have demonstrated how optical superlattices can be used to greatly enhance the capabilities of quantum gas microscopes through the measurement of the current and kinetic energy operators with local resolution and within a single experimental realization. This will enable the implementation of efficient quantum state tomography schemes~\cite{vogel_determination_1989, cramer_efficient_2010, gross_quantum_2010, lange_adaptive_2023, hearth_efficient_2023}, as well as measurements of the total energy of a quantum state and Hamiltonian learning~\cite{wiebe_hamiltonian_2014,wang_experimental_2017,yu_robust_2023,carrasco_theoretical_2021}. Furthermore, this enables to perform band structure spectroscopy to directly measure the (many-body) energy spectrum~\cite{roushan_spectroscopic_2017, xiang_simulating_2023}. The presented scheme can also find application in realizing hybrid quantum computing approaches with neutral atoms such as e.g. variational quantum algorithms~\cite{peruzzo_variational_2014,mcclean_theory_2016,michel_blueprint_2023,keijzer_pulse_2023, gonzalez-cuadra_fermionic_2023}. For this, particularly high fidelities of the rotations are important, which can be achieved for example using composite pulse sequences that cancel different inhomogeneities~\cite{wimperis_broadband_1994, cummins_tackling_2003, gevorgyan_ultrahigh-fidelity_2021}. The locally realized current measurements will also allow to detect exotic many-body states with trivial signatures in density observables, such as strongly-interacting topological phases with equilibrium currents and quantum Hall states~\cite{piraud_vortex_2015, wang_measurable_2022, kesler_single-site-resolved_2014}. 

Lastly, the possibility to perform local manipulations can be used to engineer states for the study of for example thermalization under constrained dynamics such as Hilbert space fragmentation~\cite{will_realization_2023}, lattice gauge theories~\cite{aidelsburger_cold_2022, halimeh_cold-atom_2023}, as well as to perform pair-wise entangling gates for quantum simulation~\cite{zhang_scalable_2023}. 

\section*{Acknowledgments}

The authors would like to acknowledge insightful discussions with Eugene Demler, Jens Eisert, Manuel Endres, Timon Hilker and Michael Knap. Furthermore, we would like to thank Ignacio Pérez, Scott Hubele and Sophie Häfele for technical contributions to the experimental apparatus. We received funding from the Deutsche Forschungsgemeinschaft (DFG, German Research Foundation) via Research Unit FOR5522 under project number 499180199, via Research Unit FOR 2414 under
project number 277974659 and under Germany’s Excellence Strategy – EXC-2111 – 390814868 and from the German Federal Ministry of Education and Research via the funding program quantum technologies – from basic research to market (contract number 13N15895 FermiQP). This publication has further received funding under Horizon Europe programme HORIZON-CL4-2022-QUANTUM-02-SGA via the project 101113690 (PASQuanS2.1). J.F.W. acknowledges support from the German Academic Scholarship Foundation and the Marianne-Plehn-Program.
S.H. was supported by the education and training program of the Quantum Information Research Support Center, funded through the National research foundation of Korea (NRF) by the Ministry of science and ICT (MSIT) of the Korean government (No.2021M3H3A1036573).
C.S. has received funding from the European Union’s Framework Programme for Research and Innovation Horizon 2020 (2014-2020) under the Marie Sk{\l}odowska-Curie Grant Agreement No.~754388 (LMUResearchFellows) and from LMUexcellent, funded by the BMBF and the Free State of Bavaria under the Excellence Strategy of the German Federal Government and the L\"{a}nder.


\vspace{2em}
\begin{center}
\textbf{REFERENCES}
\end{center}
\vspace{0.5em}

\putbib[manuscript]
\end{bibunit}


\clearpage

\begin{bibunit}
\setcounter{section}{0}
\setcounter{equation}{0}
\setcounter{figure}{0}
\setcounter{table}{0}
\renewcommand{\theequation}{S\arabic{equation}}
\renewcommand{\theHequation}{S\arabic{equation}}
\renewcommand{\thefigure}{S\arabic{figure}}
\renewcommand{\theHfigure}{S\arabic{figure}}
\renewcommand{\thetable}{S\arabic{table}}
\renewcommand{\theHtable}{S\arabic{table}}
\setcounter{page}{1}

\title{Local readout and control of current and kinetic energy operators in optical lattices}

\author{Alexander Impertro}
\author{Simon Karch}
\author{Julian F. Wienand}
\author{SeungJung~Huh}
\author{Christian Schweizer}
\author{Immanuel Bloch}
\author{Monika Aidelsburger}
    \affiliation{Fakult\"{a}t f\"{u}r Physik, Ludwig-Maximilians-Universit\"{a}t, 80799 Munich, Germany}
    \affiliation{Max-Planck-Institut f\"{u}r Quantenoptik, 85748 Garching, Germany}
    \affiliation{Munich Center for Quantum Science and Technology (MCQST), 80799 Munich, Germany}

\title{Supplementary Information for: \\ Local readout and control of current and kinetic energy operators in optical lattices} 

\maketitle



\section{Initial state preparation}
Our experiment begins by loading a BEC of $^{133}\rm{Cs}$ atoms into a single plane of a large-spacing vertical lattice as described in Refs.~\cite{impertro_unsupervised_2023, wienand_emergence_2023, klostermann_fast_2022}. Radial confinement is provided by a blue-detuned box potential, which we project onto the sample using a digital micromirror device (DMD) that is illuminated with incoherent light from a multi-mode laser diode ($\lambda = \SI{525}{nm}$). To lower the temperature further, we perform forced optical evaporation by exponentially decreasing the depth of the vertical lattice.

The initial states for the experiments presented in the main text are created by preparing a unity-filling Mott-insulating (MI) state in a 2D horizontal optical lattice. The $x$ axis is made with long-lattice light of wavelength $\lambda_l=\SI{1534}{nm}$, and the $y$ axis with short-lattice light at $\lambda_s=\SI{767}{nm}$. We ramp from the initial superfluid to the MI state using a two-step sigmoid ramp with a total ramp duration of \SI{400}{ms}. The final lattice depth is  $45\,E_\mathrm{r,s(l)}$ in the short(long) lattice, where $E_\mathrm{r,s(l)}=h^2/(8 m a_\mathrm{s(l)}^2)$ is the recoil energy of the short(long) lattice, $h$ is Planck's constant and $m$ is the atomic mass of cesium. During the lattice potential ramp, we increase the magnetic field to 31.5 G, tuning the scattering length from around $280\,a_0$ to approximately $500\,a_0$. After this, the short lattice potential in the $x$ direction is increased over \SI{50}{ms} to $45\,E_\mathrm{r,s}$ in presence of a finite superlattice phase offset ($\sim \pi/15$), realizing tilted double wells. Due to the tilt, all atoms localize in the lower wells, denoted as $\ket{L}$. This enables an initial-state preparation with a typical filling of $88(4)\%$ in the occupied rows and 6$(3)\%$ in the empty rows at an average imbalance of $\SI{0.88(6)}{}$, counting all sites in a region spanning $30\times36$ sites. When post-selecting on double-wells with exactly one atom, we obtain a filling of $97(2)\%$ in the occupied rows and $3(2)\%$ in the empty rows at an average imbalance of $\SI{0.93(4)}{}$.

\section{Detection}
For fluorescence imaging of $^{133}\rm{Cs}$ atoms in optical lattices, optical molasses on the D2 transition is applied, and the scattered photons are collected by a high numerical aperature (NA = 0.8) objective lens. To pin the atoms during the imaging, we abruptly ramp up the optical lattice depth to around $\SI{500}{\micro\K}$. We obtain a fluorescence image using an sCMOS camera with an exposure time of $400\,$ms. The lattice spacing is more than two times smaller than the resolution of the imaging system and in order to reconstruct the lattice occupations we employ an unsupervised deep learning algorithm that has been trained directly on our experimental images~\cite{impertro_unsupervised_2023}.

\section{Superlattice phase}
Our square superlattice consists of retro-reflected bichromatic optical lattices with wavelengths of 767\,nm and 1534\,nm. To control the superlattice phase, we utilize a beat note between the 767\,nm laser and frequency-doubled light from the 1534\,nm laser. This beat note is locked to a variable frequency offset using a phase-locked loop (Vescent D2-135), which feeds back on the frequency of the 1534\,nm laser using a piezo actuator. 

A crucial ingredient to implement $Z$ rotations is the possibility to suddenly (much faster than the dephasing time, i.e., $<\SI{1}{\milli\second}$) change from symmetric double wells ($\Delta=0$) to tilted double wells ($\Delta\gg0)$. Since the feedback loop bandwidth itself is relatively small with a rate of around $\delta\Delta/\delta T = (h\times\SI{150}{Hz}/\SI{1}{ms})$ for the lattice depths used in Fig.~\ref{fig:global_ops} of the main text, we make use of the feed-forward input of our loop controller to implement jumps faster than the feedback bandwidth. Here, an analog signal is summed onto the feedback servo output, allowing changes that are limited only by the actuator bandwidth. We calibrate the necessary feed-forward voltage by doing a slow ramp to a final tilt without any feed-forward and recording the resulting change in the servo output voltage.

\section{$\pi/2$-pulse calibration}

The pulse duration for an $X_{\pi/2}$ pulse is calibrated using bare double-well oscillations as shown in Fig.~\ref{fig:global_ops}a of the main text. Due to the finite ramp durations, during which some of the dynamics already take place, it is crucial to calibrate the pulse duration with both ramps in place (as illustrated in the left hand side of the Figure). Here, each lattice depth ramp takes $\SI{200}{\micro\second}$. We set the pulse duration to the duration for which the imbalance vanishes. Additionally, we verify this by concatenating two individual $X_{\pi/2}$ pulses, where the resulting imbalance has to match the maximum imbalance of roughly one. The $X_{\pi}$ pulse duration is calibrated equivalently, where the duration is given by the one that inverts the imbalance. Due to the finite duration of the ramps, we find that the duration of an $X_{\pi/2}$ and an $X_{\pi}$ pulse do not differ exactly by a factor of two and hence have to be independently calibrated.

\begin{figure}[!htb]
    \centering
    \includegraphics[width=\columnwidth]{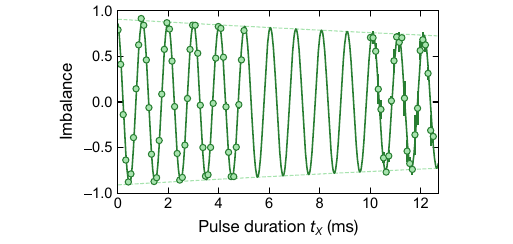}
    \caption{\textbf{$X_\pi$ pulse fidelity estimation through tunnel oscillation decay.} Bare double-well oscillations for longer overall hold times compared to Fig.~\ref{fig:global_ops}a of the main text. The solid line is a fit to an exponentially-damped sine, and the dashed lines are the exponential envelopes. The fit describes the decay of the imbalance contrast well, and we use the fitted decay constant of $\tau = \SI{57(13)}{\milli\second}$ to estimate the fidelity of a single $\pi$ pulse. The data has been evaluated in a ROI spanning $28\times 32$ sites, spatially averaging over all DWs. In addition, each data point has been averaged over several realizations (20 for the shorter pulse durations, 5 for the longer ones), and the error bars denote the standard error of the mean (s.e.m.). If invisible, they are smaller than the marker size.}
    \label{fig:sm_tunn_osc_fidelity}
\end{figure}

\section{Calibration of state preparation errors}

Due to finite temperature and imperfect initial state preparation, not all double wells are initialized in $\ket{L}$. There is a small fraction of double wells in $\ket{R}$ as well as double wells with zero or two atoms. For the data analysis, we remove the latter cases by post-selecting on double wells with exactly one atom. Here, it is crucial to identify the location of the double wells and choose the correct of the two possibilities to partition into DWs. We decide this by analyzing the intra-double well correlations ${C_{LR} = \expect{\hat{n}_L \hat{n}_R} - \expect{\hat{n}_L}\expect{\hat{n}_R}}$ as well as the fraction of removed double wells. For the wrong partitioning, we find uncorrelated wells ($C_{LR} \approx 0$) and a large fraction of removed samples, while the correct cutting gives a negative correlation ($C_{LR} \approx -0.15$) as expected and a typical fraction of removed DWs on the order of $10\%$. The wrongly-occupied DWs are not taken into account when computing average quantities. After these post-processing steps, any remaining errors stem from double-wells that are wrongly initialized in $\ket{R}$ as well as detection infidelities.

\section{$X_\pi$ pulse fidelity estimation}

To estimate the fidelity of an $X_\pi$ pulse, we record $X$ rotations for many periods in order to extract a decay envelope. As shown in Fig.~\ref{fig:sm_tunn_osc_fidelity}, the decay is well-described by a single exponential with a fitted decay constant of $\tau = \SI{57(13)}{\milli\second}$. We then define the fidelity as the ratio between the initial imbalance and the imbalance after a single $X_\pi$ pulse, which yields a fidelity of $\mathcal{F} = \SI{99.2(2)}{\percent}$. This fidelity is evaluated on a large region spanning $28\times 32$ sites, which is sensitive to large-scale potential inhomogeneities. By diminishing the size of the ROI for evaluation, even higher fidelities can in general be obtained.

\section{Disorder estimation using slow global $X$ rotations}

\begin{figure}[t]
    \centering
    \includegraphics[width=\columnwidth]{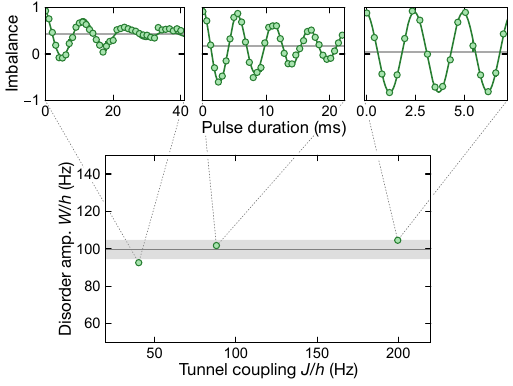}
    \caption{\textbf{Disorder estimation using global $X$ rotations.} Global $X$ rotations for three different, low-magnitude tunnel couplings. From these, we can estimate the disorder amplitude through Eq.~(\ref{eq:sm_tunn_osc_disorder}). The solid green curves in the individual  $X$ rotation traces are fits to exponentially-damped sines, and the horizontal gray lines are the extracted late-time imbalance offsets. In the bottom panel we plot the computed disorder amplitude against the fitted tunnel coupling for each trace. We extract an average disorder amplitude of $W=h\times\SI{100(5)}{Hz}$ (gray line is the average value, shaded area is the 1$\sigma$-deviation). This data has been evaluated in a ROI of $36\times40$ sites size. Error bars denote the s.e.m.}
    \label{fig:sm_tunn_osc_disorder}
\end{figure}

We estimate the on-site potential disorder in our system using global $X$ rotations at a low tunnel coupling. The results are shown in Fig.~\ref{fig:sm_tunn_osc_disorder}, where we recorded global $X$ rotations for three different tunnel couplings. For a smaller tunnel coupling, the oscillations dephase faster and become more localized toward the initial state, which is quantified by a non-zero late-time imbalance. From this steady-state imbalance offset, we can extract an estimate for the on-site potential disorder strength $W$ as
\begin{equation}
    W / J = \sqrt{\frac{6\expect{\mathcal{I}}/\mathcal{I}_0}{1-\expect{\mathcal{I}}/\mathcal{I}_0}},
    \label{eq:sm_tunn_osc_disorder}
\end{equation}
which can be derived from the single-particle dynamics in isolated DWs assuming a white-noise on-site disorder sampled from $[-W, W]$. Here, $\expect{\mathcal{I}}$ is the steady-state and $\mathcal{I}_0$ the initial imbalance, respectively. We find $W=h\times\SI{100(5)}{Hz}$, which is in good agreement with previous measurements in our experimental setup~\cite{wienand_emergence_2023}.

\section{Benchmarking of the global Ramsey and spin-echo measurements}

\subsection{Global Ramsey measurement}

The damping of imbalance oscillations during a global Ramsey measurement (Fig.~\ref{fig:global_ops}b) can be attributed to the ensemble dephasing due to random on-site potential disorder that locally changes the frequency at which a state evolves on the equator. To get an estimate for the magnitude of the on-site disorder, we assume that the on-site potential $\delta$ is randomly sampled from $\left[-W, W\right]$ (i.e. a white-noise disorder). This means that a single double well can have -- in addition to the global tilt -- a tilt between $-2W$ and $2W$. In our model we simulate the time evolution of an ensemble of 10,000 double wells, each occupied by a single particle only. The number of double wells that are initially in $\ket{L}$ and $\ket{R}$ is set to match the imbalance extracted from the $X$ rotation data $\mathcal{I}(t=0) = 0.91$ (see Fig.~\ref{fig:global_ops}b). We then fit the experimental data using this model with the disorder amplitude $W$, global tilt during the $Z$ rotation $\Delta$, and a small time offset of the imbalance oscillations as free fit parameters. The fit and the data are shown in Fig.~\ref{fig:global_ops}b of the main text, and we find $W=h\times\SI{49(2)}{Hz}$, global tilt $\Delta=h\times\SI{2.406(5)}{kHz}$ and time offset $\SI{0.349(5)}{ms}$. The value of the on-site disorder is smaller than the value estimated using tunnel oscillations (see SM Section VII.) since we perform the global Ramsey measurement in a smaller ROI with smaller on-site disorder.

\subsection{Spin-echo measurement}

To model the spin-echo measurement (Fig.~\ref{fig:global_ops}c) we follow the same numerical approach as for the global Ramsey measurement, simulating time evolution of 5,000 double wells and directly fitting the experimentally-measured data with the numerical model. The number of double wells that are initially in $\ket{L}$ and $\ket{R}$ is fixed by fitting the experimentally-measured spin-echo imbalance oscillations with a simple sine fit with Gaussian envelope and reproducing the fitted amplitude $0.663(1)$. The free fit parameters in this case are the on-site disorder amplitude $W$, global tilt $\Delta$ and a small time offset. A stronger on-site disorder leads to stronger ensamble dephasing and shorter period of rephasing during the spin-echo sequence. The fitted disorder amplitude $W=h\times\SI{53(2)}{Hz}$ is consistent with value obtained from the global Ramsey measurement.

\begin{figure}[t]
    \centering
    \includegraphics[width=\columnwidth]{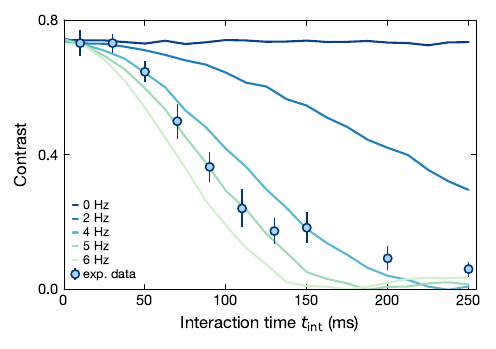}
    \caption{\textbf{Spin-echo measurement and the $T_2$ time.} Simulated spin echo contrast as a function of the interaction time $t_\mathrm{int}$. In order to explain the finite $T_2$ time, we assume additional on-site potential disorder that changes in time on a slow timescale (modelled by additional white-noise disorder sampled from $\left[ -W_\mathrm{dyn}, W_\mathrm{dyn} \right]$ in the second $Z$ pulse of the spin-echo sequence). Initial-state imbalance is chosen to match the measured contrast at $t\approx0$. Each solid line in this Figure corresponds to a different dynamical disorder amplitude $W_\mathrm{dyn}$ and is calculated using 20,000 double well samples. As expected for a spin-echo sequence, we see that in absence of additional dynamical disorder the contrast does not decrease with the total interaction time. There is a good agreement between the experimentally measured contrast and the simulation for a dynamical disorder amplitude $W_\mathrm{dyn}$ of only $h\times4-6\,\mathrm{Hz}$.}
    \label{fig:sm_spin_echo_t2_calib}
\end{figure}

In order to explain the measured $T_2$ time (Fig.~\ref{fig:global_ops}d), we assume that our on-site disorder is not entirely static but has some time-dependence. To estimate the magnitude of the time-dependent on-site disorder contribution, we add a disorder randomly sampled from interval $\left[ -W_\mathrm{dyn}, W_\mathrm{dyn} \right]$ to the second $Z$ pulse of the spin-echo sequence. Using 20,000 double well samples and the same method as for the experimental data to extract the $T_2$ time (see caption of Fig.~\ref{fig:global_ops}d), we find that a dynamical contribution to the on-site disorder with magnitude $W_\mathrm{dyn}\approx h\times 4 - 6\,\mathrm{Hz}$ can explain the observed data (see Fig.~\ref{fig:sm_spin_echo_t2_calib}).

\section{Sequence and calibration of the mask alignment for local manipulation}

\begin{figure}[t]
    \centering
    \includegraphics[width=\columnwidth]{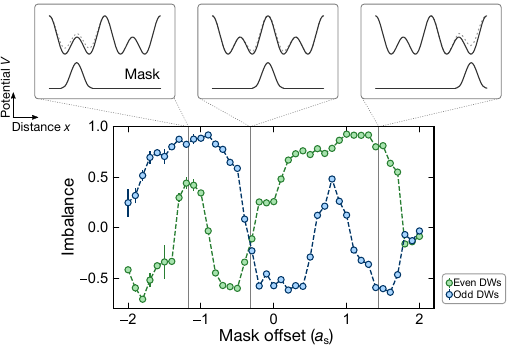}
    \caption{\textbf{Scan of the mask offset relative to the lattice.} Imbalance in the even and odd DWs as a function of the shift of the mask relative to the lattice, varied over one full period of the mask along the direction of the double wells. The underlying sequence is based on locally-detuned double-well oscillations, where the pulse duration is fixed corresponding to a $\pi$ pulse in the bare double wells. The scan shows four optimal points (i.e. where the imbalance difference between even and odd DWs is the largest), corresponding to the mask being aligned with either of the four lattice sites. The zoom-in plots show the alignment of the mask with respect to the double-well structure for three selected points.}
    \label{fig:sm_dmd_mask_shift}
\end{figure}

We use a digital micromirror device (DMD, Texas Instruments DLP6500, interface by bbs Bild- und Lichtsysteme GmbH) illuminated by laser light coming from a multimode laser diode ($\lambda=\SI{525}{nm}$ and spectral width of several nanometers) to project programmable light potentials. Apart from projecting a blue-detuned box potential, we can also locally tilt selected double wells. Doing this with high fidelity and homogeneity across larger system sizes requires a good alignment between the mask and the lattice. We fix rotation and spatial scaling using an affine transformation that maps pixels on the DMD to lattice sites in the recorded images, which we calibrate by projecting tweezers using the DMD and observing where the tweezers lie in the lattice. However, this method does not capture a shift of the mask with respect to the lattice, which, as a consequence, needs to be regularly calibrated as the lattice drifts relative to the objective (on the scale of 30 minutes to hours depending on drifts of environmental quantities). The calibration is necessary in the direction along the double wells when the mask is translationally invariant in the perpendicular direction (e.g. Fig.~\ref{fig:local_ops}c,d of the main text), or in both perpendicular directions for 2D manipulations (e.g. Fig.~\ref{fig:local_ops}e,f of the main text).

To calibrate the shift, we use locally detuned tunnel oscillations and fix the pulse duration to the first maximum in the bare double wells (e.g. around $\SI{0.4}{ms}$ in the blue trace of Fig.~\ref{fig:local_ops}a of the main text). In particular, after preparing the initial state with all DWs initialized in $\ket{L}$, we change the mask that is shown on the DMD from the box potential used during evaporation to the local tilt mask. The dynamics are then initiated by rapidly lowering the short lattice in $x$ to a depth of $12\,E_\mathrm{r,s}$ over \SI{200}{\micro\second}, and at the same time the power of the light that illuminates the DMD is raised to the power that corresponds to the target local tilt. We then wait for a given amount of time (i.e. around $\SI{0.4}{ms}$ for the mask calibration) and afterwards freeze by raising the short lattice back to $45\,E_\mathrm{r,s}$ and lowering the DMD illumination light to zero (\SI{200}{\micro\second} ramp duration).

We then scan the shift, optimizing on a maximum imbalance difference between bare and detuned double wells. An exemplary scan of the shift over one full period of the mask (4 lattice sites) is shown in Fig.~\ref{fig:sm_dmd_mask_shift}. There are four optimal points within one period, corresponding to an alignment of the mask with each one of the four different wells. Which one can in principle be chosen freely according to the application, and to keep the mask offset calibrated it is sufficient to record a few points around the chosen optimum and use them for centering.

The shift scan can in principle also be used to extract information about the resolution of the DMD imaging system. In particular, the edge steepness as well as the width of the different regions in the mask scan (e.g. the minimum in the odd DW trace around zero offset) directly depends on the resolution. It is difficult to quantify an exact number for our case, as the different alignment possibilities have a varying width and edge steepness. However, to observe the broad minimum in the odd DW trace around zero offset the resolution needs to be comparable to around one short-lattice spacing $a_\mathrm{s}$ or better.

\section{Derivation of the double-well dynamics}

To derive the dynamics after projecting into isolated DWs without making any assumptions on the (many-body) state before projection, we time-evolve the number operator under the DW Hamiltonian using the Heisenberg equations of motion. 

For an $X$ rotation, the DW Hamiltonian reads ${\hat{\mathcal{H}}_\mathrm{DW} = -J\hat{\sigma}_x}$. Under this Hamiltonian, the number difference operator evolves according to
\begin{align}
    \hat{n}_R(t) - \hat{n}_L(t) = &\left[\hat{n}_R(0) - \hat{n}_L(0)\right] \cos\left(2 J t / \hbar\right) \\ &+ i\left(\hat{a}_R^\dagger \hat{a}^{\phantom\dagger}_L - \hat{a}_L^\dagger \hat{a}^{\phantom\dagger}_R\right) \sin\left(2 J t / \hbar\right).
\end{align}
Choosing an evolution time of $\Tilde{t}=h/(8J)$, i.e., a $\pi/2$ pulse, causes the term proportional to the initial density difference to drop out, and we obtain:
\begin{align}
    \hat{n}_R(\Tilde{t}) - \hat{n}_L(\Tilde{t}) = i\left(\hat{a}_R^\dagger \hat{a}^{\phantom\dagger}_L - \hat{a}_L^\dagger \hat{a}^{\phantom\dagger}_R\right),
    \label{eq:sm_number_diff_x_pi_half}
\end{align}
which is exactly the current operator
\begin{align}
    \hat{j} = i J \left(\hat{a}_R^\dagger \hat{a}^{\phantom\dagger}_L - \hat{a}_L^\dagger \hat{a}^{\phantom\dagger}_R\right).
\end{align}
Similarly, we can compute the time evolution of the ladder operators under a $Z$ rotation, described by the Hamiltonian $\hat{\mathcal{H}}_\mathrm{DW} = -\frac{\Delta}{2}\hat{\sigma}_z$. We find
\begin{align}
    \hat{a}_R^\dagger (t) &= \hat{a}_R^\dagger (0) e^{i \Delta t / \hbar} \\
    \hat{a}_L^\dagger (t) &= \hat{a}_L^\dagger (0),
\end{align}
and, in particular, for a $Z_{\pi/2}$ rotation
\begin{align}
    \hat{a}_R^\dagger (\Tilde{t}) = i \hat{a}_R^\dagger (0),\quad \hat{a}_L^\dagger (\Tilde{t}) = \hat{a}_L^\dagger (0).
\end{align}
We insert this result into Eq.~(\ref{eq:sm_number_diff_x_pi_half}), which represents a ($Z_{\pi/2}$, $X_{\pi/2}$) pulse sequence, and find
\begin{align}
    \hat{n}_R(\Tilde{t}) - \hat{n}_L(\Tilde{t}) = - \left(\hat{a}_R^\dagger \hat{a}^{\phantom\dagger}_L + \hat{a}_L^\dagger \hat{a}^{\phantom\dagger}_R\right).
\end{align}
As postulated, this corresponds to the local kinetic energy operator $\hat{T} = -J\left(\hat{a}_R^\dagger \hat{a}^{\phantom\dagger}_L + \hat{a}_L^\dagger \hat{a}^{\phantom\dagger}_R\right)$.

\section{Laser-assisted tunneling}
We implement a laser-assisted tunneling scheme using a pair of nearly perpendicular 1534\,nm laser beams on top of the superlattice potential as originally implemented in Ref.~\cite{aidelsburger_experimental_2011}. The superlattice phase is tuned to realize tilted double wells with a large energy offset $\Delta\gg J$, such that bare tunneling is inhibited. The frequency difference of the two 1534\,nm lasers is set to satisfy the Raman condition, $\omega_1-\omega_2 = \Delta/\hbar$, which induces laser-assisted tunneling. This scheme realizes a tunnel coupling with a spatially-dependent phase.

We fit the experimentally recorded 2D correlation pattern in Fig.~\ref{fig:running_wave} of the main text to the expected correlation pattern that is parameterized by an amplitude as well as two angles. The location of a lattice site with 2D index $(m,n)$ can be written as
\begin{align}
    \vec{R} = m a_\mathrm{s} \Vec{e}_x + n a_\mathrm{s} \Vec{e}_y,
\end{align}
where $a_\mathrm{s}$ is the lattice constant of the short-period lattice and $\Vec{e}_{x,y}$ are the unit vectors along the two axes. Additionally, we express the wave vectors of the running-wave modulation beams as
\begin{align}
    \Vec{k}_1 &= k_R \left[ \cos\left(\theta_c-\theta_r\right)\Vec{e}_x + \sin\left(\theta_c-\theta_r\right)\Vec{e}_y \right] \\
    \Vec{k}_2 &= k_R \left[ -\sin\left(\theta_c+\theta_r\right)\Vec{e}_x + \cos\left(\theta_c+\theta_r\right)\Vec{e}_y \right],
\end{align}
where $k_R=\pi/(2a_\mathrm{s})$ is the magnitude of the running-wave wave vector, $\theta_c$ is a common-mode angle and $\theta_r$ a relative angle between the running-wave beams. This allows us to capture angular deviations from optimal alignment in the fit. The phase distribution $\phi_{(m,n)}$ is then given by
\begin{align}
    \phi_{(m,n)} = \left(\Vec{k}_1-\Vec{k}_2\right) \cdot \vec{R} + \phi_0,
    \label{eq:supp_phasedist}
\end{align}
where $\phi_0$ is a global phase offset that is random within $[0, 2\pi)$ in every shot. We fit the experimentally measured 2D correlator to the 2D connected correlator of the sine of Eq.~(\ref{eq:supp_phasedist}) multiplied with an amplitude factor. Additionally, we exclude the auto-correlator for the fit, as it is significantly stronger in the experiment, owing to an imperfect state preparation. The fit to the experimental data (see Fig.~\ref{fig:running_wave}c of the main text) yields an amplitude of $A=0.777(3)$, a common-mode angle of $\theta_c=\SI{1.802(3)}{^\circ}$ as well as a relative angle of $\theta_r=-\SI{0.558(8)}{^\circ}$, where the uncertainties are the standard errors of the fit. This method enables a very precise determination of the relative angles $\theta_r$ and $\theta_c$ and, in turn, to optimize the beam pointing and homogenize the phase distribution. Note however that the flux per plaquette is not significantly affected by the Moiré pattern, as it depends only on the phase difference between adjacent bonds in the $y$ direction.

\begin{center}
\textbf{SUPPLEMENTARY REFERENCES}
\end{center}
\vspace{0.5em}

\putbib[manuscript]
\end{bibunit}


\begin{thebibliography}{49}%
\makeatletter
\providecommand \@ifxundefined [1]{%
 \@ifx{#1\undefined}
}%
\providecommand \@ifnum [1]{%
 \ifnum #1\expandafter \@firstoftwo
 \else \expandafter \@secondoftwo
 \fi
}%
\providecommand \@ifx [1]{%
 \ifx #1\expandafter \@firstoftwo
 \else \expandafter \@secondoftwo
 \fi
}%
\providecommand \natexlab [1]{#1}%
\providecommand \enquote  [1]{``#1''}%
\providecommand \bibnamefont  [1]{#1}%
\providecommand \bibfnamefont [1]{#1}%
\providecommand \citenamefont [1]{#1}%
\providecommand \href@noop [0]{\@secondoftwo}%
\providecommand \href [0]{\begingroup \@sanitize@url \@href}%
\providecommand \@href[1]{\@@startlink{#1}\@@href}%
\providecommand \@@href[1]{\endgroup#1\@@endlink}%
\providecommand \@sanitize@url [0]{\catcode `\\12\catcode `\$12\catcode `\&12\catcode `\#12\catcode `\^12\catcode `\_12\catcode `\%12\relax}%
\providecommand \@@startlink[1]{}%
\providecommand \@@endlink[0]{}%
\providecommand \url  [0]{\begingroup\@sanitize@url \@url }%
\providecommand \@url [1]{\endgroup\@href {#1}{\urlprefix }}%
\providecommand \urlprefix  [0]{URL }%
\providecommand \Eprint [0]{\href }%
\providecommand \doibase [0]{https://doi.org/}%
\providecommand \selectlanguage [0]{\@gobble}%
\providecommand \bibinfo  [0]{\@secondoftwo}%
\providecommand \bibfield  [0]{\@secondoftwo}%
\providecommand \translation [1]{[#1]}%
\providecommand \BibitemOpen [0]{}%
\providecommand \bibitemStop [0]{}%
\providecommand \bibitemNoStop [0]{.\EOS\space}%
\providecommand \EOS [0]{\spacefactor3000\relax}%
\providecommand \BibitemShut  [1]{\csname bibitem#1\endcsname}%
\let\auto@bib@innerbib\@empty
\bibitem [{\citenamefont {Cirac}\ and\ \citenamefont {Zoller}(2012)}]{cirac_goals_2012}%
  \BibitemOpen
  \bibfield  {author} {\bibinfo {author} {\bibfnamefont {J.~I.}\ \bibnamefont {Cirac}}\ and\ \bibinfo {author} {\bibfnamefont {P.}~\bibnamefont {Zoller}},\ }\href {https://doi.org/10.1038/nphys2275} {\bibfield  {journal} {\bibinfo  {journal} {Nat. Phys.}\ }\textbf {\bibinfo {volume} {8}},\ \bibinfo {pages} {264} (\bibinfo {year} {2012})}\BibitemShut {NoStop}%
\bibitem [{\citenamefont {Preskill}(2018)}]{preskill_quantum_2018}%
  \BibitemOpen
  \bibfield  {author} {\bibinfo {author} {\bibfnamefont {J.}~\bibnamefont {Preskill}},\ }\href {https://doi.org/10.22331/q-2018-08-06-79} {\bibfield  {journal} {\bibinfo  {journal} {Quantum}\ }\textbf {\bibinfo {volume} {2}},\ \bibinfo {pages} {79} (\bibinfo {year} {2018})}\BibitemShut {NoStop}%
\bibitem [{\citenamefont {Trivedi}\ \emph {et~al.}(2022)\citenamefont {Trivedi}, \citenamefont {Rubio},\ and\ \citenamefont {Cirac}}]{trivedi_quantum_2022}%
  \BibitemOpen
  \bibfield  {author} {\bibinfo {author} {\bibfnamefont {R.}~\bibnamefont {Trivedi}}, \bibinfo {author} {\bibfnamefont {A.~F.}\ \bibnamefont {Rubio}},\ and\ \bibinfo {author} {\bibfnamefont {J.~I.}\ \bibnamefont {Cirac}},\ }\href {http://arxiv.org/abs/2212.04924} {\bibfield  {journal} {\bibinfo  {journal} {arXiv:2212.04924}\ } (\bibinfo {year} {2022})}\BibitemShut {NoStop}%
\bibitem [{\citenamefont {Daley}\ \emph {et~al.}(2022)\citenamefont {Daley}, \citenamefont {Bloch}, \citenamefont {Kokail}, \citenamefont {Flannigan}, \citenamefont {Pearson}, \citenamefont {Troyer},\ and\ \citenamefont {Zoller}}]{daley_practical_2022}%
  \BibitemOpen
  \bibfield  {author} {\bibinfo {author} {\bibfnamefont {A.~J.}\ \bibnamefont {Daley}}, \bibinfo {author} {\bibfnamefont {I.}~\bibnamefont {Bloch}}, \bibinfo {author} {\bibfnamefont {C.}~\bibnamefont {Kokail}}, \bibinfo {author} {\bibfnamefont {S.}~\bibnamefont {Flannigan}}, \bibinfo {author} {\bibfnamefont {N.}~\bibnamefont {Pearson}}, \bibinfo {author} {\bibfnamefont {M.}~\bibnamefont {Troyer}},\ and\ \bibinfo {author} {\bibfnamefont {P.}~\bibnamefont {Zoller}},\ }\href {https://doi.org/10.1038/s41586-022-04940-6} {\bibfield  {journal} {\bibinfo  {journal} {Nature}\ }\textbf {\bibinfo {volume} {607}},\ \bibinfo {pages} {667} (\bibinfo {year} {2022})}\BibitemShut {NoStop}%
\bibitem [{\citenamefont {Gross}\ and\ \citenamefont {Bloch}(2017)}]{gross_quantum_2017}%
  \BibitemOpen
  \bibfield  {author} {\bibinfo {author} {\bibfnamefont {C.}~\bibnamefont {Gross}}\ and\ \bibinfo {author} {\bibfnamefont {I.}~\bibnamefont {Bloch}},\ }\href {https://doi.org/10.1126/science.aal3837} {\bibfield  {journal} {\bibinfo  {journal} {Science}\ }\textbf {\bibinfo {volume} {357}},\ \bibinfo {pages} {995} (\bibinfo {year} {2017})}\BibitemShut {NoStop}%
\bibitem [{\citenamefont {Halimeh}\ \emph {et~al.}(2023)\citenamefont {Halimeh}, \citenamefont {Aidelsburger}, \citenamefont {Grusdt}, \citenamefont {Hauke},\ and\ \citenamefont {Yang}}]{halimeh_cold-atom_2023}%
  \BibitemOpen
  \bibfield  {author} {\bibinfo {author} {\bibfnamefont {J.~C.}\ \bibnamefont {Halimeh}}, \bibinfo {author} {\bibfnamefont {M.}~\bibnamefont {Aidelsburger}}, \bibinfo {author} {\bibfnamefont {F.}~\bibnamefont {Grusdt}}, \bibinfo {author} {\bibfnamefont {P.}~\bibnamefont {Hauke}},\ and\ \bibinfo {author} {\bibfnamefont {B.}~\bibnamefont {Yang}},\ }\href {http://arxiv.org/abs/2310.12201} {\bibfield  {journal} {\bibinfo  {journal} {arXiv:2310.12201}\ } (\bibinfo {year} {2023})}\BibitemShut {NoStop}%
\bibitem [{\citenamefont {Bakr}\ \emph {et~al.}(2009)\citenamefont {Bakr}, \citenamefont {Gillen}, \citenamefont {Peng}, \citenamefont {Fölling},\ and\ \citenamefont {Greiner}}]{bakr_quantum_2009}%
  \BibitemOpen
  \bibfield  {author} {\bibinfo {author} {\bibfnamefont {W.~S.}\ \bibnamefont {Bakr}}, \bibinfo {author} {\bibfnamefont {J.~I.}\ \bibnamefont {Gillen}}, \bibinfo {author} {\bibfnamefont {A.}~\bibnamefont {Peng}}, \bibinfo {author} {\bibfnamefont {S.}~\bibnamefont {Fölling}},\ and\ \bibinfo {author} {\bibfnamefont {M.}~\bibnamefont {Greiner}},\ }\href {https://doi.org/10.1038/nature08482} {\bibfield  {journal} {\bibinfo  {journal} {Nature}\ }\textbf {\bibinfo {volume} {462}},\ \bibinfo {pages} {74} (\bibinfo {year} {2009})}\BibitemShut {NoStop}%
\bibitem [{\citenamefont {Sherson}\ \emph {et~al.}(2010)\citenamefont {Sherson}, \citenamefont {Weitenberg}, \citenamefont {Endres}, \citenamefont {Cheneau}, \citenamefont {Bloch},\ and\ \citenamefont {Kuhr}}]{sherson_single-atom-resolved_2010}%
  \BibitemOpen
  \bibfield  {author} {\bibinfo {author} {\bibfnamefont {J.~F.}\ \bibnamefont {Sherson}}, \bibinfo {author} {\bibfnamefont {C.}~\bibnamefont {Weitenberg}}, \bibinfo {author} {\bibfnamefont {M.}~\bibnamefont {Endres}}, \bibinfo {author} {\bibfnamefont {M.}~\bibnamefont {Cheneau}}, \bibinfo {author} {\bibfnamefont {I.}~\bibnamefont {Bloch}},\ and\ \bibinfo {author} {\bibfnamefont {S.}~\bibnamefont {Kuhr}},\ }\href {https://doi.org/10.1038/nature09378} {\bibfield  {journal} {\bibinfo  {journal} {Nature}\ }\textbf {\bibinfo {volume} {467}},\ \bibinfo {pages} {68} (\bibinfo {year} {2010})}\BibitemShut {NoStop}%
\bibitem [{\citenamefont {Gross}\ and\ \citenamefont {Bakr}(2021)}]{gross_quantum_2021}%
  \BibitemOpen
  \bibfield  {author} {\bibinfo {author} {\bibfnamefont {C.}~\bibnamefont {Gross}}\ and\ \bibinfo {author} {\bibfnamefont {W.~S.}\ \bibnamefont {Bakr}},\ }\href {https://doi.org/10.1038/s41567-021-01370-5} {\bibfield  {journal} {\bibinfo  {journal} {Nat. Phys.}\ }\textbf {\bibinfo {volume} {17}},\ \bibinfo {pages} {1316} (\bibinfo {year} {2021})}\BibitemShut {NoStop}%
\bibitem [{\citenamefont {Swingle}\ \emph {et~al.}(2016)\citenamefont {Swingle}, \citenamefont {Bentsen}, \citenamefont {Schleier-Smith},\ and\ \citenamefont {Hayden}}]{swingle_measuring_2016}%
  \BibitemOpen
  \bibfield  {author} {\bibinfo {author} {\bibfnamefont {B.}~\bibnamefont {Swingle}}, \bibinfo {author} {\bibfnamefont {G.}~\bibnamefont {Bentsen}}, \bibinfo {author} {\bibfnamefont {M.}~\bibnamefont {Schleier-Smith}},\ and\ \bibinfo {author} {\bibfnamefont {P.}~\bibnamefont {Hayden}},\ }\href {https://doi.org/10.1103/PhysRevA.94.040302} {\bibfield  {journal} {\bibinfo  {journal} {Phys. Rev. A}\ }\textbf {\bibinfo {volume} {94}},\ \bibinfo {pages} {040302} (\bibinfo {year} {2016})}\BibitemShut {NoStop}%
\bibitem [{\citenamefont {Bohrdt}\ \emph {et~al.}(2017)\citenamefont {Bohrdt}, \citenamefont {Mendl}, \citenamefont {Endres},\ and\ \citenamefont {Knap}}]{bohrdt_scrambling_2017}%
  \BibitemOpen
  \bibfield  {author} {\bibinfo {author} {\bibfnamefont {A.}~\bibnamefont {Bohrdt}}, \bibinfo {author} {\bibfnamefont {C.~B.}\ \bibnamefont {Mendl}}, \bibinfo {author} {\bibfnamefont {M.}~\bibnamefont {Endres}},\ and\ \bibinfo {author} {\bibfnamefont {M.}~\bibnamefont {Knap}},\ }\href {https://doi.org/10.1088/1367-2630/aa719b} {\bibfield  {journal} {\bibinfo  {journal} {New J. Phys.}\ }\textbf {\bibinfo {volume} {19}},\ \bibinfo {pages} {063001} (\bibinfo {year} {2017})}\BibitemShut {NoStop}%
\bibitem [{\citenamefont {Piraud}\ \emph {et~al.}(2015)\citenamefont {Piraud}, \citenamefont {Heidrich-Meisner}, \citenamefont {McCulloch}, \citenamefont {Greschner}, \citenamefont {Vekua},\ and\ \citenamefont {Schollwöck}}]{piraud_vortex_2015}%
  \BibitemOpen
  \bibfield  {author} {\bibinfo {author} {\bibfnamefont {M.}~\bibnamefont {Piraud}}, \bibinfo {author} {\bibfnamefont {F.}~\bibnamefont {Heidrich-Meisner}}, \bibinfo {author} {\bibfnamefont {I.~P.}\ \bibnamefont {McCulloch}}, \bibinfo {author} {\bibfnamefont {S.}~\bibnamefont {Greschner}}, \bibinfo {author} {\bibfnamefont {T.}~\bibnamefont {Vekua}},\ and\ \bibinfo {author} {\bibfnamefont {U.}~\bibnamefont {Schollwöck}},\ }\href {https://doi.org/10.1103/PhysRevB.91.140406} {\bibfield  {journal} {\bibinfo  {journal} {Phys. Rev. B}\ }\textbf {\bibinfo {volume} {91}},\ \bibinfo {pages} {140406} (\bibinfo {year} {2015})}\BibitemShut {NoStop}%
\bibitem [{\citenamefont {Wang}\ \emph {et~al.}(2022)\citenamefont {Wang}, \citenamefont {Dong},\ and\ \citenamefont {Eckardt}}]{wang_measurable_2022}%
  \BibitemOpen
  \bibfield  {author} {\bibinfo {author} {\bibfnamefont {B.}~\bibnamefont {Wang}}, \bibinfo {author} {\bibfnamefont {X.}~\bibnamefont {Dong}},\ and\ \bibinfo {author} {\bibfnamefont {A.}~\bibnamefont {Eckardt}},\ }\href {https://doi.org/10.21468/SciPostPhys.12.3.095} {\bibfield  {journal} {\bibinfo  {journal} {SciPost Phys.}\ }\textbf {\bibinfo {volume} {12}},\ \bibinfo {pages} {095} (\bibinfo {year} {2022})}\BibitemShut {NoStop}%
\bibitem [{\citenamefont {Wiebe}\ \emph {et~al.}(2014)\citenamefont {Wiebe}, \citenamefont {Granade}, \citenamefont {Ferrie},\ and\ \citenamefont {Cory}}]{wiebe_hamiltonian_2014}%
  \BibitemOpen
  \bibfield  {author} {\bibinfo {author} {\bibfnamefont {N.}~\bibnamefont {Wiebe}}, \bibinfo {author} {\bibfnamefont {C.}~\bibnamefont {Granade}}, \bibinfo {author} {\bibfnamefont {C.}~\bibnamefont {Ferrie}},\ and\ \bibinfo {author} {\bibfnamefont {D.}~\bibnamefont {Cory}},\ }\href {https://doi.org/10.1103/PhysRevLett.112.190501} {\bibfield  {journal} {\bibinfo  {journal} {Phys. Rev. Lett.}\ }\textbf {\bibinfo {volume} {112}},\ \bibinfo {pages} {190501} (\bibinfo {year} {2014})}\BibitemShut {NoStop}%
\bibitem [{\citenamefont {Wang}\ \emph {et~al.}(2017)\citenamefont {Wang}, \citenamefont {Paesani}, \citenamefont {Santagati}, \citenamefont {Knauer}, \citenamefont {Gentile}, \citenamefont {Wiebe}, \citenamefont {Petruzzella}, \citenamefont {O’Brien}, \citenamefont {Rarity}, \citenamefont {Laing},\ and\ \citenamefont {Thompson}}]{wang_experimental_2017}%
  \BibitemOpen
  \bibfield  {author} {\bibinfo {author} {\bibfnamefont {J.}~\bibnamefont {Wang}}, \bibinfo {author} {\bibfnamefont {S.}~\bibnamefont {Paesani}}, \bibinfo {author} {\bibfnamefont {R.}~\bibnamefont {Santagati}}, \bibinfo {author} {\bibfnamefont {S.}~\bibnamefont {Knauer}}, \bibinfo {author} {\bibfnamefont {A.~A.}\ \bibnamefont {Gentile}}, \bibinfo {author} {\bibfnamefont {N.}~\bibnamefont {Wiebe}}, \bibinfo {author} {\bibfnamefont {M.}~\bibnamefont {Petruzzella}}, \bibinfo {author} {\bibfnamefont {J.~L.}\ \bibnamefont {O’Brien}}, \bibinfo {author} {\bibfnamefont {J.~G.}\ \bibnamefont {Rarity}}, \bibinfo {author} {\bibfnamefont {A.}~\bibnamefont {Laing}},\ and\ \bibinfo {author} {\bibfnamefont {M.~G.}\ \bibnamefont {Thompson}},\ }\href {https://doi.org/10.1038/nphys4074} {\bibfield  {journal} {\bibinfo  {journal} {Nat. Phys.}\ }\textbf {\bibinfo {volume} {13}},\ \bibinfo {pages} {551} (\bibinfo {year} {2017})}\BibitemShut {NoStop}%
\bibitem [{\citenamefont {Carrasco}\ \emph {et~al.}(2021)\citenamefont {Carrasco}, \citenamefont {Elben}, \citenamefont {Kokail}, \citenamefont {Kraus},\ and\ \citenamefont {Zoller}}]{carrasco_theoretical_2021}%
  \BibitemOpen
  \bibfield  {author} {\bibinfo {author} {\bibfnamefont {J.}~\bibnamefont {Carrasco}}, \bibinfo {author} {\bibfnamefont {A.}~\bibnamefont {Elben}}, \bibinfo {author} {\bibfnamefont {C.}~\bibnamefont {Kokail}}, \bibinfo {author} {\bibfnamefont {B.}~\bibnamefont {Kraus}},\ and\ \bibinfo {author} {\bibfnamefont {P.}~\bibnamefont {Zoller}},\ }\href {https://doi.org/10.1103/PRXQuantum.2.010102} {\bibfield  {journal} {\bibinfo  {journal} {PRX Quantum}\ }\textbf {\bibinfo {volume} {2}},\ \bibinfo {pages} {010102} (\bibinfo {year} {2021})}\BibitemShut {NoStop}%
\bibitem [{\citenamefont {Yu}\ \emph {et~al.}(2023)\citenamefont {Yu}, \citenamefont {Sun}, \citenamefont {Han},\ and\ \citenamefont {Yuan}}]{yu_robust_2023}%
  \BibitemOpen
  \bibfield  {author} {\bibinfo {author} {\bibfnamefont {W.}~\bibnamefont {Yu}}, \bibinfo {author} {\bibfnamefont {J.}~\bibnamefont {Sun}}, \bibinfo {author} {\bibfnamefont {Z.}~\bibnamefont {Han}},\ and\ \bibinfo {author} {\bibfnamefont {X.}~\bibnamefont {Yuan}},\ }\href {https://doi.org/10.22331/q-2023-06-29-1045} {\bibfield  {journal} {\bibinfo  {journal} {Quantum}\ }\textbf {\bibinfo {volume} {7}},\ \bibinfo {pages} {1045} (\bibinfo {year} {2023})}\BibitemShut {NoStop}%
\bibitem [{\citenamefont {Trotzky}\ \emph {et~al.}(2008)\citenamefont {Trotzky}, \citenamefont {Cheinet}, \citenamefont {Folling}, \citenamefont {Feld}, \citenamefont {Schnorrberger}, \citenamefont {Rey}, \citenamefont {Polkovnikov}, \citenamefont {Demler}, \citenamefont {Lukin},\ and\ \citenamefont {Bloch}}]{trotzky_time-resolved_2008}%
  \BibitemOpen
  \bibfield  {author} {\bibinfo {author} {\bibfnamefont {S.}~\bibnamefont {Trotzky}}, \bibinfo {author} {\bibfnamefont {P.}~\bibnamefont {Cheinet}}, \bibinfo {author} {\bibfnamefont {S.}~\bibnamefont {Folling}}, \bibinfo {author} {\bibfnamefont {M.}~\bibnamefont {Feld}}, \bibinfo {author} {\bibfnamefont {U.}~\bibnamefont {Schnorrberger}}, \bibinfo {author} {\bibfnamefont {A.~M.}\ \bibnamefont {Rey}}, \bibinfo {author} {\bibfnamefont {A.}~\bibnamefont {Polkovnikov}}, \bibinfo {author} {\bibfnamefont {E.~A.}\ \bibnamefont {Demler}}, \bibinfo {author} {\bibfnamefont {M.~D.}\ \bibnamefont {Lukin}},\ and\ \bibinfo {author} {\bibfnamefont {I.}~\bibnamefont {Bloch}},\ }\href {https://doi.org/10.1126/science.1150841} {\bibfield  {journal} {\bibinfo  {journal} {Science}\ }\textbf {\bibinfo {volume} {319}},\ \bibinfo {pages} {295} (\bibinfo {year} {2008})}\BibitemShut {NoStop}%
\bibitem [{\citenamefont {Trotzky}\ \emph {et~al.}(2010)\citenamefont {Trotzky}, \citenamefont {Chen}, \citenamefont {Schnorrberger}, \citenamefont {Cheinet},\ and\ \citenamefont {Bloch}}]{trotzky_controlling_2010}%
  \BibitemOpen
  \bibfield  {author} {\bibinfo {author} {\bibfnamefont {S.}~\bibnamefont {Trotzky}}, \bibinfo {author} {\bibfnamefont {Y.-A.}\ \bibnamefont {Chen}}, \bibinfo {author} {\bibfnamefont {U.}~\bibnamefont {Schnorrberger}}, \bibinfo {author} {\bibfnamefont {P.}~\bibnamefont {Cheinet}},\ and\ \bibinfo {author} {\bibfnamefont {I.}~\bibnamefont {Bloch}},\ }\href {https://doi.org/10.1103/PhysRevLett.105.265303} {\bibfield  {journal} {\bibinfo  {journal} {Phys. Rev. Lett.}\ }\textbf {\bibinfo {volume} {105}},\ \bibinfo {pages} {265303} (\bibinfo {year} {2010})}\BibitemShut {NoStop}%
\bibitem [{\citenamefont {Yang}\ \emph {et~al.}(2020)\citenamefont {Yang}, \citenamefont {Sun}, \citenamefont {Huang}, \citenamefont {Wang}, \citenamefont {Deng}, \citenamefont {Dai}, \citenamefont {Yuan},\ and\ \citenamefont {Pan}}]{yang_cooling_2020}%
  \BibitemOpen
  \bibfield  {author} {\bibinfo {author} {\bibfnamefont {B.}~\bibnamefont {Yang}}, \bibinfo {author} {\bibfnamefont {H.}~\bibnamefont {Sun}}, \bibinfo {author} {\bibfnamefont {C.-J.}\ \bibnamefont {Huang}}, \bibinfo {author} {\bibfnamefont {H.-Y.}\ \bibnamefont {Wang}}, \bibinfo {author} {\bibfnamefont {Y.}~\bibnamefont {Deng}}, \bibinfo {author} {\bibfnamefont {H.-N.}\ \bibnamefont {Dai}}, \bibinfo {author} {\bibfnamefont {Z.-S.}\ \bibnamefont {Yuan}},\ and\ \bibinfo {author} {\bibfnamefont {J.-W.}\ \bibnamefont {Pan}},\ }\href {https://doi.org/10.1126/science.aaz6801} {\bibfield  {journal} {\bibinfo  {journal} {Science}\ }\textbf {\bibinfo {volume} {369}},\ \bibinfo {pages} {550} (\bibinfo {year} {2020})}\BibitemShut {NoStop}%
\bibitem [{\citenamefont {Zhang}\ \emph {et~al.}(2023)\citenamefont {Zhang}, \citenamefont {He}, \citenamefont {Sun}, \citenamefont {Zheng}, \citenamefont {Liu}, \citenamefont {Luo}, \citenamefont {Wang}, \citenamefont {Zhu}, \citenamefont {Qiu}, \citenamefont {Shen}, \citenamefont {Wang}, \citenamefont {Lin}, \citenamefont {Yu}, \citenamefont {Li}, \citenamefont {Xiao}, \citenamefont {Li}, \citenamefont {Yang}, \citenamefont {Jiang}, \citenamefont {Dai}, \citenamefont {Zhou}, \citenamefont {Ma}, \citenamefont {Yuan},\ and\ \citenamefont {Pan}}]{zhang_scalable_2023}%
  \BibitemOpen
  \bibfield  {author} {\bibinfo {author} {\bibfnamefont {W.-Y.}\ \bibnamefont {Zhang}}, \bibinfo {author} {\bibfnamefont {M.-G.}\ \bibnamefont {He}}, \bibinfo {author} {\bibfnamefont {H.}~\bibnamefont {Sun}}, \bibinfo {author} {\bibfnamefont {Y.-G.}\ \bibnamefont {Zheng}}, \bibinfo {author} {\bibfnamefont {Y.}~\bibnamefont {Liu}}, \bibinfo {author} {\bibfnamefont {A.}~\bibnamefont {Luo}}, \bibinfo {author} {\bibfnamefont {H.-Y.}\ \bibnamefont {Wang}}, \bibinfo {author} {\bibfnamefont {Z.-H.}\ \bibnamefont {Zhu}}, \bibinfo {author} {\bibfnamefont {P.-Y.}\ \bibnamefont {Qiu}}, \bibinfo {author} {\bibfnamefont {Y.-C.}\ \bibnamefont {Shen}}, \bibinfo {author} {\bibfnamefont {X.-K.}\ \bibnamefont {Wang}}, \bibinfo {author} {\bibfnamefont {W.}~\bibnamefont {Lin}}, \bibinfo {author} {\bibfnamefont {S.-T.}\ \bibnamefont {Yu}}, \bibinfo {author} {\bibfnamefont {B.-C.}\ \bibnamefont {Li}}, \bibinfo {author} {\bibfnamefont {B.}~\bibnamefont {Xiao}}, \bibinfo {author} {\bibfnamefont {M.-D.}\ \bibnamefont {Li}}, \bibinfo
  {author} {\bibfnamefont {Y.-M.}\ \bibnamefont {Yang}}, \bibinfo {author} {\bibfnamefont {X.}~\bibnamefont {Jiang}}, \bibinfo {author} {\bibfnamefont {H.-N.}\ \bibnamefont {Dai}}, \bibinfo {author} {\bibfnamefont {Y.}~\bibnamefont {Zhou}}, \bibinfo {author} {\bibfnamefont {X.}~\bibnamefont {Ma}}, \bibinfo {author} {\bibfnamefont {Z.-S.}\ \bibnamefont {Yuan}},\ and\ \bibinfo {author} {\bibfnamefont {J.-W.}\ \bibnamefont {Pan}},\ }\href {https://doi.org/10.1103/PhysRevLett.131.073401} {\bibfield  {journal} {\bibinfo  {journal} {Phys. Rev. Lett.}\ }\textbf {\bibinfo {volume} {131}},\ \bibinfo {pages} {073401} (\bibinfo {year} {2023})}\BibitemShut {NoStop}%
\bibitem [{\citenamefont {Atala}\ \emph {et~al.}(2014)\citenamefont {Atala}, \citenamefont {Aidelsburger}, \citenamefont {Lohse}, \citenamefont {Barreiro}, \citenamefont {Paredes},\ and\ \citenamefont {Bloch}}]{atala_observation_2014}%
  \BibitemOpen
  \bibfield  {author} {\bibinfo {author} {\bibfnamefont {M.}~\bibnamefont {Atala}}, \bibinfo {author} {\bibfnamefont {M.}~\bibnamefont {Aidelsburger}}, \bibinfo {author} {\bibfnamefont {M.}~\bibnamefont {Lohse}}, \bibinfo {author} {\bibfnamefont {J.~T.}\ \bibnamefont {Barreiro}}, \bibinfo {author} {\bibfnamefont {B.}~\bibnamefont {Paredes}},\ and\ \bibinfo {author} {\bibfnamefont {I.}~\bibnamefont {Bloch}},\ }\href {https://doi.org/10.1038/nphys2998} {\bibfield  {journal} {\bibinfo  {journal} {Nat. Phys.}\ }\textbf {\bibinfo {volume} {10}},\ \bibinfo {pages} {588} (\bibinfo {year} {2014})}\BibitemShut {NoStop}%
\bibitem [{\citenamefont {Keßler}\ and\ \citenamefont {Marquardt}(2014)}]{kesler_single-site-resolved_2014}%
  \BibitemOpen
  \bibfield  {author} {\bibinfo {author} {\bibfnamefont {S.}~\bibnamefont {Keßler}}\ and\ \bibinfo {author} {\bibfnamefont {F.}~\bibnamefont {Marquardt}},\ }\href {https://doi.org/10.1103/PhysRevA.89.061601} {\bibfield  {journal} {\bibinfo  {journal} {Phys. Rev. A}\ }\textbf {\bibinfo {volume} {89}},\ \bibinfo {pages} {061601} (\bibinfo {year} {2014})}\BibitemShut {NoStop}%
\bibitem [{\citenamefont {Kaufman}\ \emph {et~al.}(2016)\citenamefont {Kaufman}, \citenamefont {Tai}, \citenamefont {Lukin}, \citenamefont {Rispoli}, \citenamefont {Schittko}, \citenamefont {Preiss},\ and\ \citenamefont {Greiner}}]{kaufman_quantum_2016}%
  \BibitemOpen
  \bibfield  {author} {\bibinfo {author} {\bibfnamefont {A.~M.}\ \bibnamefont {Kaufman}}, \bibinfo {author} {\bibfnamefont {M.~E.}\ \bibnamefont {Tai}}, \bibinfo {author} {\bibfnamefont {A.}~\bibnamefont {Lukin}}, \bibinfo {author} {\bibfnamefont {M.}~\bibnamefont {Rispoli}}, \bibinfo {author} {\bibfnamefont {R.}~\bibnamefont {Schittko}}, \bibinfo {author} {\bibfnamefont {P.~M.}\ \bibnamefont {Preiss}},\ and\ \bibinfo {author} {\bibfnamefont {M.}~\bibnamefont {Greiner}},\ }\href {https://doi.org/10.1126/science.aaf6725} {\bibfield  {journal} {\bibinfo  {journal} {Science}\ }\textbf {\bibinfo {volume} {353}},\ \bibinfo {pages} {794} (\bibinfo {year} {2016})}\BibitemShut {NoStop}%
\bibitem [{\citenamefont {Su}\ \emph {et~al.}(2023)\citenamefont {Su}, \citenamefont {Sun}, \citenamefont {Hudomal}, \citenamefont {Desaules}, \citenamefont {Zhou}, \citenamefont {Yang}, \citenamefont {Halimeh}, \citenamefont {Yuan}, \citenamefont {Papić},\ and\ \citenamefont {Pan}}]{su_observation_2023}%
  \BibitemOpen
  \bibfield  {author} {\bibinfo {author} {\bibfnamefont {G.-X.}\ \bibnamefont {Su}}, \bibinfo {author} {\bibfnamefont {H.}~\bibnamefont {Sun}}, \bibinfo {author} {\bibfnamefont {A.}~\bibnamefont {Hudomal}}, \bibinfo {author} {\bibfnamefont {J.-Y.}\ \bibnamefont {Desaules}}, \bibinfo {author} {\bibfnamefont {Z.-Y.}\ \bibnamefont {Zhou}}, \bibinfo {author} {\bibfnamefont {B.}~\bibnamefont {Yang}}, \bibinfo {author} {\bibfnamefont {J.~C.}\ \bibnamefont {Halimeh}}, \bibinfo {author} {\bibfnamefont {Z.-S.}\ \bibnamefont {Yuan}}, \bibinfo {author} {\bibfnamefont {Z.}~\bibnamefont {Papić}},\ and\ \bibinfo {author} {\bibfnamefont {J.-W.}\ \bibnamefont {Pan}},\ }\href {https://doi.org/10.1103/PhysRevResearch.5.023010} {\bibfield  {journal} {\bibinfo  {journal} {Phys. Rev. Res.}\ }\textbf {\bibinfo {volume} {5}},\ \bibinfo {pages} {023010} (\bibinfo {year} {2023})}\BibitemShut {NoStop}%
\bibitem [{\citenamefont {Fölling}\ \emph {et~al.}(2007)\citenamefont {Fölling}, \citenamefont {Trotzky}, \citenamefont {Cheinet}, \citenamefont {Feld}, \citenamefont {Saers}, \citenamefont {Widera}, \citenamefont {Müller},\ and\ \citenamefont {Bloch}}]{folling_direct_2007}%
  \BibitemOpen
  \bibfield  {author} {\bibinfo {author} {\bibfnamefont {S.}~\bibnamefont {Fölling}}, \bibinfo {author} {\bibfnamefont {S.}~\bibnamefont {Trotzky}}, \bibinfo {author} {\bibfnamefont {P.}~\bibnamefont {Cheinet}}, \bibinfo {author} {\bibfnamefont {M.}~\bibnamefont {Feld}}, \bibinfo {author} {\bibfnamefont {R.}~\bibnamefont {Saers}}, \bibinfo {author} {\bibfnamefont {A.}~\bibnamefont {Widera}}, \bibinfo {author} {\bibfnamefont {T.}~\bibnamefont {Müller}},\ and\ \bibinfo {author} {\bibfnamefont {I.}~\bibnamefont {Bloch}},\ }\href {https://doi.org/10.1038/nature06112} {\bibfield  {journal} {\bibinfo  {journal} {Nature}\ }\textbf {\bibinfo {volume} {448}},\ \bibinfo {pages} {1029} (\bibinfo {year} {2007})}\BibitemShut {NoStop}%
\bibitem [{\citenamefont {Impertro}\ \emph {et~al.}(2023)\citenamefont {Impertro}, \citenamefont {Wienand}, \citenamefont {Häfele}, \citenamefont {von Raven}, \citenamefont {Hubele}, \citenamefont {Klostermann}, \citenamefont {Cabrera}, \citenamefont {Bloch},\ and\ \citenamefont {Aidelsburger}}]{impertro_unsupervised_2023}%
  \BibitemOpen
  \bibfield  {author} {\bibinfo {author} {\bibfnamefont {A.}~\bibnamefont {Impertro}}, \bibinfo {author} {\bibfnamefont {J.~F.}\ \bibnamefont {Wienand}}, \bibinfo {author} {\bibfnamefont {S.}~\bibnamefont {Häfele}}, \bibinfo {author} {\bibfnamefont {H.}~\bibnamefont {von Raven}}, \bibinfo {author} {\bibfnamefont {S.}~\bibnamefont {Hubele}}, \bibinfo {author} {\bibfnamefont {T.}~\bibnamefont {Klostermann}}, \bibinfo {author} {\bibfnamefont {C.~R.}\ \bibnamefont {Cabrera}}, \bibinfo {author} {\bibfnamefont {I.}~\bibnamefont {Bloch}},\ and\ \bibinfo {author} {\bibfnamefont {M.}~\bibnamefont {Aidelsburger}},\ }\href {https://doi.org/10.1038/s42005-023-01287-w} {\bibfield  {journal} {\bibinfo  {journal} {Commun. Phys.}\ }\textbf {\bibinfo {volume} {6}},\ \bibinfo {pages} {1} (\bibinfo {year} {2023})}\BibitemShut {NoStop}%
\bibitem [{\citenamefont {Wienand}\ \emph {et~al.}(2023)\citenamefont {Wienand}, \citenamefont {Karch}, \citenamefont {Impertro}, \citenamefont {Schweizer}, \citenamefont {McCulloch}, \citenamefont {Vasseur}, \citenamefont {Gopalakrishnan}, \citenamefont {Aidelsburger},\ and\ \citenamefont {Bloch}}]{wienand_emergence_2023}%
  \BibitemOpen
  \bibfield  {author} {\bibinfo {author} {\bibfnamefont {J.~F.}\ \bibnamefont {Wienand}}, \bibinfo {author} {\bibfnamefont {S.}~\bibnamefont {Karch}}, \bibinfo {author} {\bibfnamefont {A.}~\bibnamefont {Impertro}}, \bibinfo {author} {\bibfnamefont {C.}~\bibnamefont {Schweizer}}, \bibinfo {author} {\bibfnamefont {E.}~\bibnamefont {McCulloch}}, \bibinfo {author} {\bibfnamefont {R.}~\bibnamefont {Vasseur}}, \bibinfo {author} {\bibfnamefont {S.}~\bibnamefont {Gopalakrishnan}}, \bibinfo {author} {\bibfnamefont {M.}~\bibnamefont {Aidelsburger}},\ and\ \bibinfo {author} {\bibfnamefont {I.}~\bibnamefont {Bloch}},\ }\href {http://arxiv.org/abs/2306.11457} {\bibfield  {journal} {\bibinfo  {journal} {arXiv.2306.11457}\ } (\bibinfo {year} {2023})}\BibitemShut {NoStop}%
\bibitem [{\citenamefont {Shaw}\ \emph {et~al.}(2023)\citenamefont {Shaw}, \citenamefont {Finkelstein}, \citenamefont {Tsai}, \citenamefont {Scholl}, \citenamefont {Yoon}, \citenamefont {Choi},\ and\ \citenamefont {Endres}}]{shaw_multi-ensemble_2023}%
  \BibitemOpen
  \bibfield  {author} {\bibinfo {author} {\bibfnamefont {A.~L.}\ \bibnamefont {Shaw}}, \bibinfo {author} {\bibfnamefont {R.}~\bibnamefont {Finkelstein}}, \bibinfo {author} {\bibfnamefont {R.~B.-S.}\ \bibnamefont {Tsai}}, \bibinfo {author} {\bibfnamefont {P.}~\bibnamefont {Scholl}}, \bibinfo {author} {\bibfnamefont {T.~H.}\ \bibnamefont {Yoon}}, \bibinfo {author} {\bibfnamefont {J.}~\bibnamefont {Choi}},\ and\ \bibinfo {author} {\bibfnamefont {M.}~\bibnamefont {Endres}},\ }\href {http://arxiv.org/abs/2303.16885} {\bibfield  {journal} {\bibinfo  {journal} {arXiv:2303.16885}\ } (\bibinfo {year} {2023})}\BibitemShut {NoStop}%
\bibitem [{\citenamefont {Aidelsburger}\ \emph {et~al.}(2011)\citenamefont {Aidelsburger}, \citenamefont {Atala}, \citenamefont {Nascimbène}, \citenamefont {Trotzky}, \citenamefont {Chen},\ and\ \citenamefont {Bloch}}]{aidelsburger_experimental_2011}%
  \BibitemOpen
  \bibfield  {author} {\bibinfo {author} {\bibfnamefont {M.}~\bibnamefont {Aidelsburger}}, \bibinfo {author} {\bibfnamefont {M.}~\bibnamefont {Atala}}, \bibinfo {author} {\bibfnamefont {S.}~\bibnamefont {Nascimbène}}, \bibinfo {author} {\bibfnamefont {S.}~\bibnamefont {Trotzky}}, \bibinfo {author} {\bibfnamefont {Y.-A.}\ \bibnamefont {Chen}},\ and\ \bibinfo {author} {\bibfnamefont {I.}~\bibnamefont {Bloch}},\ }\href {https://doi.org/10.1103/PhysRevLett.107.255301} {\bibfield  {journal} {\bibinfo  {journal} {Phys. Rev. Lett.}\ }\textbf {\bibinfo {volume} {107}},\ \bibinfo {pages} {255301} (\bibinfo {year} {2011})}\BibitemShut {NoStop}%
\bibitem [{\citenamefont {Miyake}\ \emph {et~al.}(2013)\citenamefont {Miyake}, \citenamefont {Siviloglou}, \citenamefont {Kennedy}, \citenamefont {Burton},\ and\ \citenamefont {Ketterle}}]{miyake_realizing_2013}%
  \BibitemOpen
  \bibfield  {author} {\bibinfo {author} {\bibfnamefont {H.}~\bibnamefont {Miyake}}, \bibinfo {author} {\bibfnamefont {G.~A.}\ \bibnamefont {Siviloglou}}, \bibinfo {author} {\bibfnamefont {C.~J.}\ \bibnamefont {Kennedy}}, \bibinfo {author} {\bibfnamefont {W.~C.}\ \bibnamefont {Burton}},\ and\ \bibinfo {author} {\bibfnamefont {W.}~\bibnamefont {Ketterle}},\ }\href {https://doi.org/10.1103/PhysRevLett.111.185302} {\bibfield  {journal} {\bibinfo  {journal} {Phys. Rev. Lett.}\ }\textbf {\bibinfo {volume} {111}},\ \bibinfo {pages} {185302} (\bibinfo {year} {2013})}\BibitemShut {NoStop}%
\bibitem [{\citenamefont {Tai}\ \emph {et~al.}(2017)\citenamefont {Tai}, \citenamefont {Lukin}, \citenamefont {Rispoli}, \citenamefont {Schittko}, \citenamefont {Menke}, \citenamefont {Borgnia}, \citenamefont {Preiss}, \citenamefont {Grusdt}, \citenamefont {Kaufman},\ and\ \citenamefont {Greiner}}]{tai_microscopy_2017}%
  \BibitemOpen
  \bibfield  {author} {\bibinfo {author} {\bibfnamefont {M.~E.}\ \bibnamefont {Tai}}, \bibinfo {author} {\bibfnamefont {A.}~\bibnamefont {Lukin}}, \bibinfo {author} {\bibfnamefont {M.}~\bibnamefont {Rispoli}}, \bibinfo {author} {\bibfnamefont {R.}~\bibnamefont {Schittko}}, \bibinfo {author} {\bibfnamefont {T.}~\bibnamefont {Menke}}, \bibinfo {author} {\bibfnamefont {D.}~\bibnamefont {Borgnia}}, \bibinfo {author} {\bibfnamefont {P.~M.}\ \bibnamefont {Preiss}}, \bibinfo {author} {\bibfnamefont {F.}~\bibnamefont {Grusdt}}, \bibinfo {author} {\bibfnamefont {A.~M.}\ \bibnamefont {Kaufman}},\ and\ \bibinfo {author} {\bibfnamefont {M.}~\bibnamefont {Greiner}},\ }\href {https://doi.org/10.1038/nature22811} {\bibfield  {journal} {\bibinfo  {journal} {Nature}\ }\textbf {\bibinfo {volume} {546}},\ \bibinfo {pages} {519} (\bibinfo {year} {2017})}\BibitemShut {NoStop}%
\bibitem [{\citenamefont {Vogel}\ and\ \citenamefont {Risken}(1989)}]{vogel_determination_1989}%
  \BibitemOpen
  \bibfield  {author} {\bibinfo {author} {\bibfnamefont {K.}~\bibnamefont {Vogel}}\ and\ \bibinfo {author} {\bibfnamefont {H.}~\bibnamefont {Risken}},\ }\href {https://doi.org/10.1103/PhysRevA.40.2847} {\bibfield  {journal} {\bibinfo  {journal} {Phys. Rev. A}\ }\textbf {\bibinfo {volume} {40}},\ \bibinfo {pages} {2847} (\bibinfo {year} {1989})}\BibitemShut {NoStop}%
\bibitem [{\citenamefont {Cramer}\ \emph {et~al.}(2010)\citenamefont {Cramer}, \citenamefont {Plenio}, \citenamefont {Flammia}, \citenamefont {Somma}, \citenamefont {Gross}, \citenamefont {Bartlett}, \citenamefont {Landon-Cardinal}, \citenamefont {Poulin},\ and\ \citenamefont {Liu}}]{cramer_efficient_2010}%
  \BibitemOpen
  \bibfield  {author} {\bibinfo {author} {\bibfnamefont {M.}~\bibnamefont {Cramer}}, \bibinfo {author} {\bibfnamefont {M.~B.}\ \bibnamefont {Plenio}}, \bibinfo {author} {\bibfnamefont {S.~T.}\ \bibnamefont {Flammia}}, \bibinfo {author} {\bibfnamefont {R.}~\bibnamefont {Somma}}, \bibinfo {author} {\bibfnamefont {D.}~\bibnamefont {Gross}}, \bibinfo {author} {\bibfnamefont {S.~D.}\ \bibnamefont {Bartlett}}, \bibinfo {author} {\bibfnamefont {O.}~\bibnamefont {Landon-Cardinal}}, \bibinfo {author} {\bibfnamefont {D.}~\bibnamefont {Poulin}},\ and\ \bibinfo {author} {\bibfnamefont {Y.-K.}\ \bibnamefont {Liu}},\ }\href {https://doi.org/10.1038/ncomms1147} {\bibfield  {journal} {\bibinfo  {journal} {Nat. Commun.}\ }\textbf {\bibinfo {volume} {1}},\ \bibinfo {pages} {149} (\bibinfo {year} {2010})}\BibitemShut {NoStop}%
\bibitem [{\citenamefont {Gross}\ \emph {et~al.}(2010)\citenamefont {Gross}, \citenamefont {Liu}, \citenamefont {Flammia}, \citenamefont {Becker},\ and\ \citenamefont {Eisert}}]{gross_quantum_2010}%
  \BibitemOpen
  \bibfield  {author} {\bibinfo {author} {\bibfnamefont {D.}~\bibnamefont {Gross}}, \bibinfo {author} {\bibfnamefont {Y.-K.}\ \bibnamefont {Liu}}, \bibinfo {author} {\bibfnamefont {S.~T.}\ \bibnamefont {Flammia}}, \bibinfo {author} {\bibfnamefont {S.}~\bibnamefont {Becker}},\ and\ \bibinfo {author} {\bibfnamefont {J.}~\bibnamefont {Eisert}},\ }\href {https://doi.org/10.1103/PhysRevLett.105.150401} {\bibfield  {journal} {\bibinfo  {journal} {Phys. Rev. Lett.}\ }\textbf {\bibinfo {volume} {105}},\ \bibinfo {pages} {150401} (\bibinfo {year} {2010})}\BibitemShut {NoStop}%
\bibitem [{\citenamefont {Lange}\ \emph {et~al.}(2023)\citenamefont {Lange}, \citenamefont {Kebrič}, \citenamefont {Buser}, \citenamefont {Schollwöck}, \citenamefont {Grusdt},\ and\ \citenamefont {Bohrdt}}]{lange_adaptive_2023}%
  \BibitemOpen
  \bibfield  {author} {\bibinfo {author} {\bibfnamefont {H.}~\bibnamefont {Lange}}, \bibinfo {author} {\bibfnamefont {M.}~\bibnamefont {Kebrič}}, \bibinfo {author} {\bibfnamefont {M.}~\bibnamefont {Buser}}, \bibinfo {author} {\bibfnamefont {U.}~\bibnamefont {Schollwöck}}, \bibinfo {author} {\bibfnamefont {F.}~\bibnamefont {Grusdt}},\ and\ \bibinfo {author} {\bibfnamefont {A.}~\bibnamefont {Bohrdt}},\ }\href {https://doi.org/10.22331/q-2023-10-09-1129} {\bibfield  {journal} {\bibinfo  {journal} {Quantum}\ }\textbf {\bibinfo {volume} {7}},\ \bibinfo {pages} {1129} (\bibinfo {year} {2023})}\BibitemShut {NoStop}%
\bibitem [{\citenamefont {Hearth}\ \emph {et~al.}(2023)\citenamefont {Hearth}, \citenamefont {Flynn}, \citenamefont {Chandran},\ and\ \citenamefont {Laumann}}]{hearth_efficient_2023}%
  \BibitemOpen
  \bibfield  {author} {\bibinfo {author} {\bibfnamefont {S.~N.}\ \bibnamefont {Hearth}}, \bibinfo {author} {\bibfnamefont {M.~O.}\ \bibnamefont {Flynn}}, \bibinfo {author} {\bibfnamefont {A.}~\bibnamefont {Chandran}},\ and\ \bibinfo {author} {\bibfnamefont {C.~R.}\ \bibnamefont {Laumann}},\ }\href {http://arxiv.org/abs/2311.09291} {\bibfield  {journal} {\bibinfo  {journal} {arXiv.2311.09291}\ } (\bibinfo {year} {2023})}\BibitemShut {NoStop}%
\bibitem [{\citenamefont {Roushan}\ \emph {et~al.}(2017)\citenamefont {Roushan}, \citenamefont {Neill}, \citenamefont {Tangpanitanon}, \citenamefont {Bastidas}, \citenamefont {Megrant}, \citenamefont {Barends}, \citenamefont {Chen}, \citenamefont {Chen}, \citenamefont {Chiaro}, \citenamefont {Dunsworth}, \citenamefont {Fowler}, \citenamefont {Foxen}, \citenamefont {Giustina}, \citenamefont {Jeffrey}, \citenamefont {Kelly}, \citenamefont {Lucero}, \citenamefont {Mutus}, \citenamefont {Neeley}, \citenamefont {Quintana}, \citenamefont {Sank}, \citenamefont {Vainsencher}, \citenamefont {Wenner}, \citenamefont {White}, \citenamefont {Neven}, \citenamefont {Angelakis},\ and\ \citenamefont {Martinis}}]{roushan_spectroscopic_2017}%
  \BibitemOpen
  \bibfield  {author} {\bibinfo {author} {\bibfnamefont {P.}~\bibnamefont {Roushan}}, \bibinfo {author} {\bibfnamefont {C.}~\bibnamefont {Neill}}, \bibinfo {author} {\bibfnamefont {J.}~\bibnamefont {Tangpanitanon}}, \bibinfo {author} {\bibfnamefont {V.~M.}\ \bibnamefont {Bastidas}}, \bibinfo {author} {\bibfnamefont {A.}~\bibnamefont {Megrant}}, \bibinfo {author} {\bibfnamefont {R.}~\bibnamefont {Barends}}, \bibinfo {author} {\bibfnamefont {Y.}~\bibnamefont {Chen}}, \bibinfo {author} {\bibfnamefont {Z.}~\bibnamefont {Chen}}, \bibinfo {author} {\bibfnamefont {B.}~\bibnamefont {Chiaro}}, \bibinfo {author} {\bibfnamefont {A.}~\bibnamefont {Dunsworth}}, \bibinfo {author} {\bibfnamefont {A.}~\bibnamefont {Fowler}}, \bibinfo {author} {\bibfnamefont {B.}~\bibnamefont {Foxen}}, \bibinfo {author} {\bibfnamefont {M.}~\bibnamefont {Giustina}}, \bibinfo {author} {\bibfnamefont {E.}~\bibnamefont {Jeffrey}}, \bibinfo {author} {\bibfnamefont {J.}~\bibnamefont {Kelly}}, \bibinfo {author} {\bibfnamefont {E.}~\bibnamefont
  {Lucero}}, \bibinfo {author} {\bibfnamefont {J.}~\bibnamefont {Mutus}}, \bibinfo {author} {\bibfnamefont {M.}~\bibnamefont {Neeley}}, \bibinfo {author} {\bibfnamefont {C.}~\bibnamefont {Quintana}}, \bibinfo {author} {\bibfnamefont {D.}~\bibnamefont {Sank}}, \bibinfo {author} {\bibfnamefont {A.}~\bibnamefont {Vainsencher}}, \bibinfo {author} {\bibfnamefont {J.}~\bibnamefont {Wenner}}, \bibinfo {author} {\bibfnamefont {T.}~\bibnamefont {White}}, \bibinfo {author} {\bibfnamefont {H.}~\bibnamefont {Neven}}, \bibinfo {author} {\bibfnamefont {D.~G.}\ \bibnamefont {Angelakis}},\ and\ \bibinfo {author} {\bibfnamefont {J.}~\bibnamefont {Martinis}},\ }\href {https://doi.org/10.1126/science.aao1401} {\bibfield  {journal} {\bibinfo  {journal} {Science}\ }\textbf {\bibinfo {volume} {358}},\ \bibinfo {pages} {1175} (\bibinfo {year} {2017})}\BibitemShut {NoStop}%
\bibitem [{\citenamefont {Xiang}\ \emph {et~al.}(2023)\citenamefont {Xiang}, \citenamefont {Huang}, \citenamefont {Zhang}, \citenamefont {Liu}, \citenamefont {Shi}, \citenamefont {Deng}, \citenamefont {Liu}, \citenamefont {Li}, \citenamefont {Liang}, \citenamefont {Mei}, \citenamefont {Yu}, \citenamefont {Xue}, \citenamefont {Tian}, \citenamefont {Song}, \citenamefont {Liu}, \citenamefont {Xu}, \citenamefont {Zheng}, \citenamefont {Nori},\ and\ \citenamefont {Fan}}]{xiang_simulating_2023}%
  \BibitemOpen
  \bibfield  {author} {\bibinfo {author} {\bibfnamefont {Z.-C.}\ \bibnamefont {Xiang}}, \bibinfo {author} {\bibfnamefont {K.}~\bibnamefont {Huang}}, \bibinfo {author} {\bibfnamefont {Y.-R.}\ \bibnamefont {Zhang}}, \bibinfo {author} {\bibfnamefont {T.}~\bibnamefont {Liu}}, \bibinfo {author} {\bibfnamefont {Y.-H.}\ \bibnamefont {Shi}}, \bibinfo {author} {\bibfnamefont {C.-L.}\ \bibnamefont {Deng}}, \bibinfo {author} {\bibfnamefont {T.}~\bibnamefont {Liu}}, \bibinfo {author} {\bibfnamefont {H.}~\bibnamefont {Li}}, \bibinfo {author} {\bibfnamefont {G.-H.}\ \bibnamefont {Liang}}, \bibinfo {author} {\bibfnamefont {Z.-Y.}\ \bibnamefont {Mei}}, \bibinfo {author} {\bibfnamefont {H.}~\bibnamefont {Yu}}, \bibinfo {author} {\bibfnamefont {G.}~\bibnamefont {Xue}}, \bibinfo {author} {\bibfnamefont {Y.}~\bibnamefont {Tian}}, \bibinfo {author} {\bibfnamefont {X.}~\bibnamefont {Song}}, \bibinfo {author} {\bibfnamefont {Z.-B.}\ \bibnamefont {Liu}}, \bibinfo {author} {\bibfnamefont {K.}~\bibnamefont {Xu}}, \bibinfo {author}
  {\bibfnamefont {D.}~\bibnamefont {Zheng}}, \bibinfo {author} {\bibfnamefont {F.}~\bibnamefont {Nori}},\ and\ \bibinfo {author} {\bibfnamefont {H.}~\bibnamefont {Fan}},\ }\href {https://doi.org/10.1038/s41467-023-41230-9} {\bibfield  {journal} {\bibinfo  {journal} {Nat. Commun.}\ }\textbf {\bibinfo {volume} {14}},\ \bibinfo {pages} {5433} (\bibinfo {year} {2023})}\BibitemShut {NoStop}%
\bibitem [{\citenamefont {Peruzzo}\ \emph {et~al.}(2014)\citenamefont {Peruzzo}, \citenamefont {McClean}, \citenamefont {Shadbolt}, \citenamefont {Yung}, \citenamefont {Zhou}, \citenamefont {Love}, \citenamefont {Aspuru-Guzik},\ and\ \citenamefont {O’Brien}}]{peruzzo_variational_2014}%
  \BibitemOpen
  \bibfield  {author} {\bibinfo {author} {\bibfnamefont {A.}~\bibnamefont {Peruzzo}}, \bibinfo {author} {\bibfnamefont {J.}~\bibnamefont {McClean}}, \bibinfo {author} {\bibfnamefont {P.}~\bibnamefont {Shadbolt}}, \bibinfo {author} {\bibfnamefont {M.-H.}\ \bibnamefont {Yung}}, \bibinfo {author} {\bibfnamefont {X.-Q.}\ \bibnamefont {Zhou}}, \bibinfo {author} {\bibfnamefont {P.~J.}\ \bibnamefont {Love}}, \bibinfo {author} {\bibfnamefont {A.}~\bibnamefont {Aspuru-Guzik}},\ and\ \bibinfo {author} {\bibfnamefont {J.~L.}\ \bibnamefont {O’Brien}},\ }\href {https://doi.org/10.1038/ncomms5213} {\bibfield  {journal} {\bibinfo  {journal} {Nat. Commun.}\ }\textbf {\bibinfo {volume} {5}},\ \bibinfo {pages} {4213} (\bibinfo {year} {2014})}\BibitemShut {NoStop}%
\bibitem [{\citenamefont {McClean}\ \emph {et~al.}(2016)\citenamefont {McClean}, \citenamefont {Romero}, \citenamefont {Babbush},\ and\ \citenamefont {Aspuru-Guzik}}]{mcclean_theory_2016}%
  \BibitemOpen
  \bibfield  {author} {\bibinfo {author} {\bibfnamefont {J.~R.}\ \bibnamefont {McClean}}, \bibinfo {author} {\bibfnamefont {J.}~\bibnamefont {Romero}}, \bibinfo {author} {\bibfnamefont {R.}~\bibnamefont {Babbush}},\ and\ \bibinfo {author} {\bibfnamefont {A.}~\bibnamefont {Aspuru-Guzik}},\ }\href {https://doi.org/10.1088/1367-2630/18/2/023023} {\bibfield  {journal} {\bibinfo  {journal} {New J. Phys.}\ }\textbf {\bibinfo {volume} {18}},\ \bibinfo {pages} {023023} (\bibinfo {year} {2016})}\BibitemShut {NoStop}%
\bibitem [{\citenamefont {Michel}\ \emph {et~al.}(2023)\citenamefont {Michel}, \citenamefont {Grijalva}, \citenamefont {Henriet}, \citenamefont {Domain},\ and\ \citenamefont {Browaeys}}]{michel_blueprint_2023}%
  \BibitemOpen
  \bibfield  {author} {\bibinfo {author} {\bibfnamefont {A.}~\bibnamefont {Michel}}, \bibinfo {author} {\bibfnamefont {S.}~\bibnamefont {Grijalva}}, \bibinfo {author} {\bibfnamefont {L.}~\bibnamefont {Henriet}}, \bibinfo {author} {\bibfnamefont {C.}~\bibnamefont {Domain}},\ and\ \bibinfo {author} {\bibfnamefont {A.}~\bibnamefont {Browaeys}},\ }\href {https://doi.org/10.1103/PhysRevA.107.042602} {\bibfield  {journal} {\bibinfo  {journal} {Phys. Rev. A}\ }\textbf {\bibinfo {volume} {107}},\ \bibinfo {pages} {042602} (\bibinfo {year} {2023})}\BibitemShut {NoStop}%
\bibitem [{\citenamefont {Keijzer}\ \emph {et~al.}(2023)\citenamefont {Keijzer}, \citenamefont {Tse},\ and\ \citenamefont {Kokkelmans}}]{keijzer_pulse_2023}%
  \BibitemOpen
  \bibfield  {author} {\bibinfo {author} {\bibfnamefont {R.~d.}\ \bibnamefont {Keijzer}}, \bibinfo {author} {\bibfnamefont {O.}~\bibnamefont {Tse}},\ and\ \bibinfo {author} {\bibfnamefont {S.}~\bibnamefont {Kokkelmans}},\ }\href {https://doi.org/10.22331/q-2023-01-26-908} {\bibfield  {journal} {\bibinfo  {journal} {Quantum}\ }\textbf {\bibinfo {volume} {7}},\ \bibinfo {pages} {908} (\bibinfo {year} {2023})}\BibitemShut {NoStop}%
\bibitem [{\citenamefont {González-Cuadra}\ \emph {et~al.}(2023)\citenamefont {González-Cuadra}, \citenamefont {Bluvstein}, \citenamefont {Kalinowski}, \citenamefont {Kaubruegger}, \citenamefont {Maskara}, \citenamefont {Naldesi}, \citenamefont {Zache}, \citenamefont {Kaufman}, \citenamefont {Lukin}, \citenamefont {Pichler}, \citenamefont {Vermersch}, \citenamefont {Ye},\ and\ \citenamefont {Zoller}}]{gonzalez-cuadra_fermionic_2023}%
  \BibitemOpen
  \bibfield  {author} {\bibinfo {author} {\bibfnamefont {D.}~\bibnamefont {González-Cuadra}}, \bibinfo {author} {\bibfnamefont {D.}~\bibnamefont {Bluvstein}}, \bibinfo {author} {\bibfnamefont {M.}~\bibnamefont {Kalinowski}}, \bibinfo {author} {\bibfnamefont {R.}~\bibnamefont {Kaubruegger}}, \bibinfo {author} {\bibfnamefont {N.}~\bibnamefont {Maskara}}, \bibinfo {author} {\bibfnamefont {P.}~\bibnamefont {Naldesi}}, \bibinfo {author} {\bibfnamefont {T.~V.}\ \bibnamefont {Zache}}, \bibinfo {author} {\bibfnamefont {A.~M.}\ \bibnamefont {Kaufman}}, \bibinfo {author} {\bibfnamefont {M.~D.}\ \bibnamefont {Lukin}}, \bibinfo {author} {\bibfnamefont {H.}~\bibnamefont {Pichler}}, \bibinfo {author} {\bibfnamefont {B.}~\bibnamefont {Vermersch}}, \bibinfo {author} {\bibfnamefont {J.}~\bibnamefont {Ye}},\ and\ \bibinfo {author} {\bibfnamefont {P.}~\bibnamefont {Zoller}},\ }\href {https://doi.org/10.1073/pnas.2304294120} {\bibfield  {journal} {\bibinfo  {journal} {Proc. Natl. Acad. Sci. U.S.A.}\ }\textbf {\bibinfo {volume}
  {120}},\ \bibinfo {pages} {2304294120} (\bibinfo {year} {2023})}\BibitemShut {NoStop}%
\bibitem [{\citenamefont {Wimperis}(1994)}]{wimperis_broadband_1994}%
  \BibitemOpen
  \bibfield  {author} {\bibinfo {author} {\bibfnamefont {S.}~\bibnamefont {Wimperis}},\ }\href {https://doi.org/10.1006/jmra.1994.1159} {\bibfield  {journal} {\bibinfo  {journal} {J. magn. reson., Ser. A}\ }\textbf {\bibinfo {volume} {109}},\ \bibinfo {pages} {221} (\bibinfo {year} {1994})}\BibitemShut {NoStop}%
\bibitem [{\citenamefont {Cummins}\ \emph {et~al.}(2003)\citenamefont {Cummins}, \citenamefont {Llewellyn},\ and\ \citenamefont {Jones}}]{cummins_tackling_2003}%
  \BibitemOpen
  \bibfield  {author} {\bibinfo {author} {\bibfnamefont {H.~K.}\ \bibnamefont {Cummins}}, \bibinfo {author} {\bibfnamefont {G.}~\bibnamefont {Llewellyn}},\ and\ \bibinfo {author} {\bibfnamefont {J.~A.}\ \bibnamefont {Jones}},\ }\href {https://doi.org/10.1103/PhysRevA.67.042308} {\bibfield  {journal} {\bibinfo  {journal} {Phys. Rev. A}\ }\textbf {\bibinfo {volume} {67}},\ \bibinfo {pages} {042308} (\bibinfo {year} {2003})}\BibitemShut {NoStop}%
\bibitem [{\citenamefont {Gevorgyan}\ and\ \citenamefont {Vitanov}(2021)}]{gevorgyan_ultrahigh-fidelity_2021}%
  \BibitemOpen
  \bibfield  {author} {\bibinfo {author} {\bibfnamefont {H.~L.}\ \bibnamefont {Gevorgyan}}\ and\ \bibinfo {author} {\bibfnamefont {N.~V.}\ \bibnamefont {Vitanov}},\ }\href {https://doi.org/10.1103/PhysRevA.104.012609} {\bibfield  {journal} {\bibinfo  {journal} {Phys. Rev. A}\ }\textbf {\bibinfo {volume} {104}},\ \bibinfo {pages} {012609} (\bibinfo {year} {2021})}\BibitemShut {NoStop}%
\bibitem [{\citenamefont {Will}\ \emph {et~al.}(2023)\citenamefont {Will}, \citenamefont {Moessner},\ and\ \citenamefont {Pollmann}}]{will_realization_2023}%
  \BibitemOpen
  \bibfield  {author} {\bibinfo {author} {\bibfnamefont {M.}~\bibnamefont {Will}}, \bibinfo {author} {\bibfnamefont {R.}~\bibnamefont {Moessner}},\ and\ \bibinfo {author} {\bibfnamefont {F.}~\bibnamefont {Pollmann}},\ }\href {https://arxiv.org/abs/2311.05695} {\bibfield  {journal} {\bibinfo  {journal} {arXiv:2311.05695}\ } (\bibinfo {year} {2023})}\BibitemShut {NoStop}%
\bibitem [{\citenamefont {Aidelsburger}\ \emph {et~al.}(2022)\citenamefont {Aidelsburger}, \citenamefont {Barbiero}, \citenamefont {Bermudez}, \citenamefont {Chanda}, \citenamefont {Dauphin}, \citenamefont {González-Cuadra}, \citenamefont {Grzybowski}, \citenamefont {Hands}, \citenamefont {Jendrzejewski}, \citenamefont {Jünemann}, \citenamefont {Juzeliunas}, \citenamefont {Kasper}, \citenamefont {Piga}, \citenamefont {Ran}, \citenamefont {Rizzi}, \citenamefont {Sierra}, \citenamefont {Tagliacozzo}, \citenamefont {Tirrito}, \citenamefont {Zache}, \citenamefont {Zakrzewski}, \citenamefont {Zohar},\ and\ \citenamefont {Lewenstein}}]{aidelsburger_cold_2022}%
  \BibitemOpen
  \bibfield  {author} {\bibinfo {author} {\bibfnamefont {M.}~\bibnamefont {Aidelsburger}}, \bibinfo {author} {\bibfnamefont {L.}~\bibnamefont {Barbiero}}, \bibinfo {author} {\bibfnamefont {A.}~\bibnamefont {Bermudez}}, \bibinfo {author} {\bibfnamefont {T.}~\bibnamefont {Chanda}}, \bibinfo {author} {\bibfnamefont {A.}~\bibnamefont {Dauphin}}, \bibinfo {author} {\bibfnamefont {D.}~\bibnamefont {González-Cuadra}}, \bibinfo {author} {\bibfnamefont {P.~R.}\ \bibnamefont {Grzybowski}}, \bibinfo {author} {\bibfnamefont {S.}~\bibnamefont {Hands}}, \bibinfo {author} {\bibfnamefont {F.}~\bibnamefont {Jendrzejewski}}, \bibinfo {author} {\bibfnamefont {J.}~\bibnamefont {Jünemann}}, \bibinfo {author} {\bibfnamefont {G.}~\bibnamefont {Juzeliunas}}, \bibinfo {author} {\bibfnamefont {V.}~\bibnamefont {Kasper}}, \bibinfo {author} {\bibfnamefont {A.}~\bibnamefont {Piga}}, \bibinfo {author} {\bibfnamefont {S.-J.}\ \bibnamefont {Ran}}, \bibinfo {author} {\bibfnamefont {M.}~\bibnamefont {Rizzi}}, \bibinfo {author}
  {\bibfnamefont {G.}~\bibnamefont {Sierra}}, \bibinfo {author} {\bibfnamefont {L.}~\bibnamefont {Tagliacozzo}}, \bibinfo {author} {\bibfnamefont {E.}~\bibnamefont {Tirrito}}, \bibinfo {author} {\bibfnamefont {T.~V.}\ \bibnamefont {Zache}}, \bibinfo {author} {\bibfnamefont {J.}~\bibnamefont {Zakrzewski}}, \bibinfo {author} {\bibfnamefont {E.}~\bibnamefont {Zohar}},\ and\ \bibinfo {author} {\bibfnamefont {M.}~\bibnamefont {Lewenstein}},\ }\href {https://doi.org/10.1098/rsta.2021.0064} {\bibfield  {journal} {\bibinfo  {journal} {Phil. Trans. R. Soc. A.}\ }\textbf {\bibinfo {volume} {380}},\ \bibinfo {pages} {20210064} (\bibinfo {year} {2022})}\BibitemShut {NoStop}%
\end{thebibliography}%


\begin{thebibliography}{4}%
\makeatletter
\providecommand \@ifxundefined [1]{%
 \@ifx{#1\undefined}
}%
\providecommand \@ifnum [1]{%
 \ifnum #1\expandafter \@firstoftwo
 \else \expandafter \@secondoftwo
 \fi
}%
\providecommand \@ifx [1]{%
 \ifx #1\expandafter \@firstoftwo
 \else \expandafter \@secondoftwo
 \fi
}%
\providecommand \natexlab [1]{#1}%
\providecommand \enquote  [1]{``#1''}%
\providecommand \bibnamefont  [1]{#1}%
\providecommand \bibfnamefont [1]{#1}%
\providecommand \citenamefont [1]{#1}%
\providecommand \href@noop [0]{\@secondoftwo}%
\providecommand \href [0]{\begingroup \@sanitize@url \@href}%
\providecommand \@href[1]{\@@startlink{#1}\@@href}%
\providecommand \@@href[1]{\endgroup#1\@@endlink}%
\providecommand \@sanitize@url [0]{\catcode `\\12\catcode `\$12\catcode `\&12\catcode `\#12\catcode `\^12\catcode `\_12\catcode `\%12\relax}%
\providecommand \@@startlink[1]{}%
\providecommand \@@endlink[0]{}%
\providecommand \url  [0]{\begingroup\@sanitize@url \@url }%
\providecommand \@url [1]{\endgroup\@href {#1}{\urlprefix }}%
\providecommand \urlprefix  [0]{URL }%
\providecommand \Eprint [0]{\href }%
\providecommand \doibase [0]{https://doi.org/}%
\providecommand \selectlanguage [0]{\@gobble}%
\providecommand \bibinfo  [0]{\@secondoftwo}%
\providecommand \bibfield  [0]{\@secondoftwo}%
\providecommand \translation [1]{[#1]}%
\providecommand \BibitemOpen [0]{}%
\providecommand \bibitemStop [0]{}%
\providecommand \bibitemNoStop [0]{.\EOS\space}%
\providecommand \EOS [0]{\spacefactor3000\relax}%
\providecommand \BibitemShut  [1]{\csname bibitem#1\endcsname}%
\let\auto@bib@innerbib\@empty
\bibitem [{\citenamefont {Impertro}\ \emph {et~al.}(2023)\citenamefont {Impertro}, \citenamefont {Wienand}, \citenamefont {Häfele}, \citenamefont {von Raven}, \citenamefont {Hubele}, \citenamefont {Klostermann}, \citenamefont {Cabrera}, \citenamefont {Bloch},\ and\ \citenamefont {Aidelsburger}}]{impertro_unsupervised_2023}%
  \BibitemOpen
  \bibfield  {author} {\bibinfo {author} {\bibfnamefont {A.}~\bibnamefont {Impertro}}, \bibinfo {author} {\bibfnamefont {J.~F.}\ \bibnamefont {Wienand}}, \bibinfo {author} {\bibfnamefont {S.}~\bibnamefont {Häfele}}, \bibinfo {author} {\bibfnamefont {H.}~\bibnamefont {von Raven}}, \bibinfo {author} {\bibfnamefont {S.}~\bibnamefont {Hubele}}, \bibinfo {author} {\bibfnamefont {T.}~\bibnamefont {Klostermann}}, \bibinfo {author} {\bibfnamefont {C.~R.}\ \bibnamefont {Cabrera}}, \bibinfo {author} {\bibfnamefont {I.}~\bibnamefont {Bloch}},\ and\ \bibinfo {author} {\bibfnamefont {M.}~\bibnamefont {Aidelsburger}},\ }\href {https://doi.org/10.1038/s42005-023-01287-w} {\bibfield  {journal} {\bibinfo  {journal} {Commun. Phys.}\ }\textbf {\bibinfo {volume} {6}},\ \bibinfo {pages} {1} (\bibinfo {year} {2023})}\BibitemShut {NoStop}%
\bibitem [{\citenamefont {Wienand}\ \emph {et~al.}(2023)\citenamefont {Wienand}, \citenamefont {Karch}, \citenamefont {Impertro}, \citenamefont {Schweizer}, \citenamefont {McCulloch}, \citenamefont {Vasseur}, \citenamefont {Gopalakrishnan}, \citenamefont {Aidelsburger},\ and\ \citenamefont {Bloch}}]{wienand_emergence_2023}%
  \BibitemOpen
  \bibfield  {author} {\bibinfo {author} {\bibfnamefont {J.~F.}\ \bibnamefont {Wienand}}, \bibinfo {author} {\bibfnamefont {S.}~\bibnamefont {Karch}}, \bibinfo {author} {\bibfnamefont {A.}~\bibnamefont {Impertro}}, \bibinfo {author} {\bibfnamefont {C.}~\bibnamefont {Schweizer}}, \bibinfo {author} {\bibfnamefont {E.}~\bibnamefont {McCulloch}}, \bibinfo {author} {\bibfnamefont {R.}~\bibnamefont {Vasseur}}, \bibinfo {author} {\bibfnamefont {S.}~\bibnamefont {Gopalakrishnan}}, \bibinfo {author} {\bibfnamefont {M.}~\bibnamefont {Aidelsburger}},\ and\ \bibinfo {author} {\bibfnamefont {I.}~\bibnamefont {Bloch}},\ }\href {http://arxiv.org/abs/2306.11457} {\bibfield  {journal} {\bibinfo  {journal} {arXiv.2306.11457}\ } (\bibinfo {year} {2023})}\BibitemShut {NoStop}%
\bibitem [{\citenamefont {Klostermann}\ \emph {et~al.}(2022)\citenamefont {Klostermann}, \citenamefont {Cabrera}, \citenamefont {von Raven}, \citenamefont {Wienand}, \citenamefont {Schweizer}, \citenamefont {Bloch},\ and\ \citenamefont {Aidelsburger}}]{klostermann_fast_2022}%
  \BibitemOpen
  \bibfield  {author} {\bibinfo {author} {\bibfnamefont {T.}~\bibnamefont {Klostermann}}, \bibinfo {author} {\bibfnamefont {C.~R.}\ \bibnamefont {Cabrera}}, \bibinfo {author} {\bibfnamefont {H.}~\bibnamefont {von Raven}}, \bibinfo {author} {\bibfnamefont {J.~F.}\ \bibnamefont {Wienand}}, \bibinfo {author} {\bibfnamefont {C.}~\bibnamefont {Schweizer}}, \bibinfo {author} {\bibfnamefont {I.}~\bibnamefont {Bloch}},\ and\ \bibinfo {author} {\bibfnamefont {M.}~\bibnamefont {Aidelsburger}},\ }\href {https://doi.org/10.1103/PhysRevA.105.043319} {\bibfield  {journal} {\bibinfo  {journal} {Phys. Rev. A}\ }\textbf {\bibinfo {volume} {105}},\ \bibinfo {pages} {043319} (\bibinfo {year} {2022})}\BibitemShut {NoStop}%
\bibitem [{\citenamefont {Aidelsburger}\ \emph {et~al.}(2011)\citenamefont {Aidelsburger}, \citenamefont {Atala}, \citenamefont {Nascimbène}, \citenamefont {Trotzky}, \citenamefont {Chen},\ and\ \citenamefont {Bloch}}]{aidelsburger_experimental_2011}%
  \BibitemOpen
  \bibfield  {author} {\bibinfo {author} {\bibfnamefont {M.}~\bibnamefont {Aidelsburger}}, \bibinfo {author} {\bibfnamefont {M.}~\bibnamefont {Atala}}, \bibinfo {author} {\bibfnamefont {S.}~\bibnamefont {Nascimbène}}, \bibinfo {author} {\bibfnamefont {S.}~\bibnamefont {Trotzky}}, \bibinfo {author} {\bibfnamefont {Y.-A.}\ \bibnamefont {Chen}},\ and\ \bibinfo {author} {\bibfnamefont {I.}~\bibnamefont {Bloch}},\ }\href {https://doi.org/10.1103/PhysRevLett.107.255301} {\bibfield  {journal} {\bibinfo  {journal} {Phys. Rev. Lett.}\ }\textbf {\bibinfo {volume} {107}},\ \bibinfo {pages} {255301} (\bibinfo {year} {2011})}\BibitemShut {NoStop}%
\end{thebibliography}%


\begin{thebibliography}{0}%
\makeatletter
\providecommand \@ifxundefined [1]{%
 \@ifx{#1\undefined}
}%
\providecommand \@ifnum [1]{%
 \ifnum #1\expandafter \@firstoftwo
 \else \expandafter \@secondoftwo
 \fi
}%
\providecommand \@ifx [1]{%
 \ifx #1\expandafter \@firstoftwo
 \else \expandafter \@secondoftwo
 \fi
}%
\providecommand \natexlab [1]{#1}%
\providecommand \enquote  [1]{``#1''}%
\providecommand \bibnamefont  [1]{#1}%
\providecommand \bibfnamefont [1]{#1}%
\providecommand \citenamefont [1]{#1}%
\providecommand \href@noop [0]{\@secondoftwo}%
\providecommand \href [0]{\begingroup \@sanitize@url \@href}%
\providecommand \@href[1]{\@@startlink{#1}\@@href}%
\providecommand \@@href[1]{\endgroup#1\@@endlink}%
\providecommand \@sanitize@url [0]{\catcode `\\12\catcode `\$12\catcode `\&12\catcode `\#12\catcode `\^12\catcode `\_12\catcode `\%12\relax}%
\providecommand \@@startlink[1]{}%
\providecommand \@@endlink[0]{}%
\providecommand \url  [0]{\begingroup\@sanitize@url \@url }%
\providecommand \@url [1]{\endgroup\@href {#1}{\urlprefix }}%
\providecommand \urlprefix  [0]{URL }%
\providecommand \Eprint [0]{\href }%
\providecommand \doibase [0]{https://doi.org/}%
\providecommand \selectlanguage [0]{\@gobble}%
\providecommand \bibinfo  [0]{\@secondoftwo}%
\providecommand \bibfield  [0]{\@secondoftwo}%
\providecommand \translation [1]{[#1]}%
\providecommand \BibitemOpen [0]{}%
\providecommand \bibitemStop [0]{}%
\providecommand \bibitemNoStop [0]{.\EOS\space}%
\providecommand \EOS [0]{\spacefactor3000\relax}%
\providecommand \BibitemShut  [1]{\csname bibitem#1\endcsname}%
\let\auto@bib@innerbib\@empty
\end{thebibliography}%
\end{document}